\documentclass{aa}  
\usepackage{graphicx}
\usepackage{caption}
\usepackage{txfonts}
\usepackage[colorlinks=true,allcolors=blue,urlcolor=blue]{hyperref}

\newcommand{\teff}{T_{\mathrm{eff}}}
\newcommand{\lgg}{\log{g}}
\newcommand{\feh}{\mathrm{\left[Fe/H\right]}}
\newcommand{\dex}{\mathrm{dex}}

\usepackage{color}
\usepackage[normalem]{ulem}

\begin{document} 

   \title{Non-LTE analysis of K I in late-type stars}

   \author{Henrique Reggiani\inst{1,2}
          \and
          Anish M. Amarsi\inst{2}
          \and
          Karin Lind\inst{2,3}
          \and
                  Paul S. Barklem\inst{4}
                  \and
                  Oleg Zatsarinny\inst{5}
                  \and
                  Klaus Bartschat\inst{5}
                 \and
                 Dmitry V. Fursa\inst{6}
                 \and
                 Igor Bray\inst{6}
                 \and 
                 Lorenzo Spina\inst{7}
                 \and
                 Jorge Mel\'endez\inst{1}  
        }

   \institute{
                Universidade de S\~ao Paulo, Instituto de Astronomia, Geof\'isica e Ci\^encias Atmosf\'ericas, IAG, 
                Departamento de Astronomia, Rua do Mat\~ao 1226, Cidade Universit\'aria, 05508-900, SP, Brazil. 
        \email{hreggiani@gmail.com}
     \and
        Max-Planck Institute for Astronomy, Konigstuhl 17, 69117 Heidelberg, Germany.
     \and
        Observational Astrophysics, Department of Physics and Astronomy, Uppsala University, Box 516, SE-751 
        20 Uppsala, Sweden.
     \and
        Theoretical Astrophysics, Department of Physics and Astronomy, Uppsala University, Box 516, SE-751 
        20 Uppsala, Sweden.
    \and
        Department of Physics and Astronomy, Drake University, Des Moines, Iowa, 50311, USA.
        \email{oleg\_zoi@yahoo.com}
    \and
        Curtin Institute for Computation and Department of Physics and Astronomy, 
        Kent Street, Bentley, Perth, Western Australia 6102, Australia
    \and
        Monash Centre for Astrophysics, School of Physics and Astronomy, Monash University, VIC 3800, Australia.
        }

   \date{29/01/2019}

 
  \abstract
   {Older models of Galactic chemical evolution (GCE) predict
  [K/Fe] ratios as much as $1\,\dex$ lower than those inferred from stellar observations. 
   Abundances of potassium are mainly 
   based on analyses of the $7698\ \AA$ resonance line,
   and the discrepancy between GCE models and observations
   is in part caused by the assumption of local thermodynamic equilibrium
   (LTE) in spectroscopic analyses.}
   {We study the statistical equilibrium of \ion{K}{I}, focusing on the non-LTE effects on the $7698 \ \AA$ 
   line. We aim to determine how non-LTE abundances of potassium can improve 
   the analysis of its chemical evolution, and help to constrain the yields of GCE models. }
   {We construct a new model \ion{K}{I} atom that employs
   the most up-to-date atomic data.
   In particular, we calculate and present 
   inelastic e+K collisional excitation cross-sections
   from the convergent close-coupling (CCC) and the 
   $B$-Spline $R$-matrix (BSR) methods, and H+K collisions from the two-electron model (LCAO).
   We constructed a fine, extended grid of non-LTE abundance corrections based on 1D MARCS models that span 
   $4000<\teff / \rm{K}<8000$, $0.50<\lgg<5.00$, $-5.00<\feh<+0.50$, and applied the corrections to 
   potassium abundances extracted from the literature.
   }
   {In concordance with previous studies, we find severe non-LTE effects in the $7698 \ \AA$ line.  
   The line is stronger in non-LTE and the abundance corrections can reach $\sim-0.7\,\dex$ 
   for solar-metallicity stars such as Procyon.
   We determine potassium abundances in six benchmark stars, and obtain consistent results from different 
   optical lines. We explore the effects of atmospheric inhomogeneity by computing for the first time a full 3D 
   non-LTE stellar spectrum of \ion{K}{I} lines for a test star. We find that 3D modeling is necessary to 
   predict a correct shape of the resonance 7698Å line, but the line strength is similar to that found in 1D non-LTE. 
   }
   {Our non-LTE abundance corrections reduce the scatter and change the cosmic trends of literature 
   potassium abundances. In the regime [Fe/H]$\lesssim-1.0$ the non-LTE abundances show a good agreement 
   with the GCE model with yields from rotating massive stars. The reduced scatter 
   of the non-LTE corrected 
   abundances of a sample of solar twins shows that line-by-line differential analysis techniques cannot fully 
   compensate for systematic LTE modelling errors; the scatter introduced by such errors introduces a 
   spurious dispersion to K evolution.}
   

   \maketitle
%

\section{Introduction}

Potassium is an alkali metal with an atomic structure very similar to that of sodium (so similar that 
they were mistakenly believed to be the same element until after the eighteenth century). However, K is 
typically an order-of-magnitude less abundant than Na and its spectral fingerprint in late-type stars is 
accordingly weaker and much less studied. 
Potassium has three stable isotopes ($^{39}$K, $^{40}$K and $^{41}$K); all produced via 
hydrostatic oxygen shell burning and explosive oxygen burning in massive stars, with a relative proportion 
that depends on the stellar mass \citep{woosley1995}. The lightest isotope is dominant with 93\% occurrence 
in solar-system meteorites  \citep{lodders2009}. Because of this, and the negligible isotopic shifts of atomic 
K lines \citep{clayton2007}, to our knowledge potassium isotopic ratios have not yet been measured in stars. 

Assuming that there are no additional nucleosynthetic production sites of K, there is a clear shortage in 
the supernova yields, as evidenced by the existence of a large discrepancy between models of chemical 
evolution and observed K abundances obtained via stellar spectroscopy \citep{zhao2016,sneden2016}. 
To resolve this discrepancy, the supernova yields for K would need to be empirically increased by as much as 
twice what current theory would suggest \citep[e.g., ][]{takeda2002,romano2010}. \cite{kobayashi2011} 
speculate that the underproduction of K in the models is at least partially due to the lack of a neutrino process. 
On the other hand, yields of rotating massive stars improve the agreement between models and observations, 
especially in the metal-poor ([Fe/H]$\le -2.0$) regime, where the scatter of the observed abundances 
starts to increase and the model predictions match at least the lower envelope \citep{prantzos2018}.

The discrepancy between models of chemical evolution and observed potassium abundances may also 
in part be caused by systematic errors in modeling the main potassium abundance diagnostic, the
resonance \ion{K}{I} $7664$ and $7698$ $\AA$ doublet. In practice, heavy blends with telluric O$_2$ make 
it difficult to correctly 
assess the potassium abundances using the $7664 \ \AA$ line, meaning that most of the measurements 
of potassium come from the $7698 \ \AA$ line. Although there are two other observable 
\ion{K}{I} lines in the optical spectra ($5801$ and $6939 \AA$), these are usually weak and 
can only be measured in  cool (T$_{\rm{eff}} \lesssim 6000$ K) high-metallicity 
([Fe/H]$\sim +0.0$) stars, and are therefore not used as a diagnostic of the 
potassium abundance in most studies. 

In 1975 astronomers already knew that the $7698 \AA$ line was sensitive to departures 
from local thermodynamic equilibrium \citep[LTE;][]{reza1975,bruls1992,takeda2002,zhang2006,andrievsky2010,zhao2016}.
Previous Galactic chemical evolution (GCE) studies have demonstrated that LTE potassium abundances can be 
more than 1 dex higher than those predicted by existing models \citep{kobayashi2006,prantzos2018}. 
\citet{takeda2002} studied the departures from LTE in
the \ion{K}{I} $7698.9 \ \AA$ line across a grid of $100$ atmospheric models. They found non-LTE 
corrections spanning from $-0.2$ to $-0.7$ dex, with a strong sensitivity to 
effective temperature, which was also confirmed in later works \citep{takeda2009,andrievsky2010}.
Thus, non-LTE modeling can significantly decrease the discrepancy between models and observations 
\citep[e.g.,][]{kobayashi2006,romano2010,kobayashi2011,prantzos2018}.

Non-LTE abundances for K are computed, as is common for late-type stars, under the trace element 
assumption that neglects feedback on the atmospheric structure. Solving the statistical equilibrium 
equations requires a wealth of atomic data; in particular radiative and collisional transition probabilities. 
The main uncertainties in non-LTE analyses of potassium
have hitherto originated from the photoionization cross-sections, the inelastic H+K collisions, and 
the inelastic e+K collisions; however, the situation has recently improved. 
Potassium, atomic number 19, falls just outside of the scope of the Opacity Project 
\citep{seaton1996,badnell2005}, and previous studies have employed hydrogenic approximations. 
However, \cite{zatsarinny2010} calculated photoionization cross-sections with the
fully relativistic Dirac B-spline R-matrix (DBSR) method. Furthermore,  \cite{yakovleva2018} 
recently presented new inelastic H+K collisions using the new asymptotic two-electron model of \cite{barklem2016}; 
this recipe predicts rates that are in reasonable agreement with fully ab-initio quantum mechanical
calculations for low-excitation transitions of lithium, sodium, and magnesium, in particular for the processes with 
the largest rates.

Rates for inelastic e+K collisional excitation have, in the past, typically been estimated using the 
semi-empirical recipe of \cite{park71}, hereafter Park71, or the semi-empirical formula from 
\cite{vanreg1962}, hereafter vanReg62. Several more accurate 
methods now exist. In this study, we present new data based on two modern close-coupling methods, namely 
the convergent close coupling method (CCC) and the B-spline R-matrix method (BSR).
\cite{osorio2011} and \cite{barklem2017} demonstrated that the rate coefficients calculated 
via these two methods tend to agree better than a factor of two for lithium and magnesium.

With the goal of improving potassium abundance determinations, here we study the non-LTE effects in potassium. 
To that end, we construct a new model \ion{K}{I} atom that employs more accurate atomic data than before. 
In particular, we calculate inelastic e+K collisional excitation cross-sections from the 
CCC and BSR methods and we also employ improved photoionization cross-sections
and inelastic H+K collisions from the literature. In Sect. \ref{ecol} we present and discuss the 
calculations of the inelastic e+K collisional excitation cross-sections
from the CCC and BSR methods, and in Sect. \ref{atom} we present the atomic model.
In Sect. \ref{nlte_in_potassium} we discuss the departures from LTE, and compare results among different 
collisional and photoionization recipes. In Sect. \ref{analysis} we show the line fits of our 
non-LTE model against the solar spectrum and the abundance analysis of benchmark stars 
(HD 103095, HD 84937, HD 140283, HD 192263 and Procyon). In Sect. \ref{grid} we describe 
our non-LTE grid of corrections. In Sect. \ref{gce} we discuss the implications for Galactic
chemical evolution and we conclude in Sect. \ref{conclusao}.

\section{Inelastic e+K collisional excitation}
\label{ecol}

Calculations of collisional excitation of K by electron impacts were performed with two state-of-the-art  
close-coupling methods, the CCC and the BSR. 
These methods and calculations are described below, in Sect. \ref{calc_ccc} and Sect. \ref{calc_bsr}. 
In Sect. \ref{ek_data} the resulting data are described, and are compared in 
Sect. \ref{coll_data}, along with older calculations.

\subsection{Convergent close coupling method}
\label{calc_ccc}
The e+K CCC calculations are based on the generalization of the e-H
formalism \citep{bray1992} to quasi one-electron targets such as atomic Li, Na, and
K \citep{bray1994}. The valence electron is treated as the active electron
on top of a frozen Hartree-Fock core. Additionally, virtual excitation
of the core electrons is treated via phenomenological local
polarization potentials. Their parameters are adjusted to yield the
optimally accurate
one-electron excitation energies of the valence electron for each
target orbital angular momentum $l$.

The e+K system has been considered
previously \citep{SKBT98,SKBT99,SKBT01}, which demonstrated the applicability 
of the CCC approach to the collision system at all energies. The key
issue with the CCC method is to choose a sufficiently large number of
Laguerre-based states for convergence in the required physical quantities of interest
to the desired level of precision. Convergence considerations are
energy and transition dependent. To make the presentation simpler we
take a single large Laguerre basis $N_l$ chosen to generate
sufficiently accurate states for the
transitions and energies of interest. Specifically, we take the
maximum orbital angular momentum $l_{\rm max}=6$, and take
$N_l=40-l$. Such a choice leads to $n\le9$ physical eigenstates with
the remainder being negative- and positive-energy pseudostates. Some of the states generated have very high energies and
may be excluded from the calculations, depending on the incident
electron energy. Having chosen the Laguerre
basis, the calculations proceed as described in \cite{bray1994}.

\subsection{B-spline R-matrix method}
\label{calc_bsr}
An overview of the $B$-spline $R$-matrix (BSR), which is a different
and entirely independent implementation for solving the close-coupling equations,
can be found in \cite{BSR+}. The calculations were performed
with an extended version of the computer
code \citep{zatsarinny2006} that allows for the inclusion of a sufficient number
of physical target states as well as continuum pseudostates
in the intermediate-energy regime.  Like CCC, this  $R$-matrix with pseudo-states
implementation is expected to provide a converged (with the number of states included)
solution of the close-coupling equations, with the remaining differences between the CCC and BSR results most likely
being related to a slightly different target description.

All target states considered in the present calculations
have the quasi-one-electron structure (core)$nl$,
with the core configuration K$^+(1s^22s^22p^63s^23p^6)$.  
We started the structure-part of the problem by generating the core orbitals from a 
Hartree-Fock (HF) calculation for K$^+$. 
The principal correlation effects in the atomic states are
related to the core-valence interaction. In many
calculations for alkali-metal atoms (see also the CCC description above), a phenomenological one-electron
core polarization potential is typically added to
account for this effect. Although such a
potential simplifies the calculations significantly and can
provide accurate excitation energies and oscillator strengths,
the question still remains as to how well the model potential can
simulate the entire core-valence correlation, including nondipole
contributions. In the present BSR approach, we 
therefore chose to include the core-valence correlation {\it ab initio\/} through 
the polarized-pseudostate approach. 
This method is described in detail in our previous calculation for photoionization of potassium 
\citep{zatsarinny2010}.  

Specifically, the target states were expanded as  
\begin{equation}\label{eq:Psi}
\Psi (3p^6nl,LS)  = {\cal A}[\Phi (3p^6)P(nl)]^{LS}  \nonumber + {\cal A} \sum_{k=1}^3 [\phi _p^k P(n'l')]^{LS}, 
\end{equation}
where ${\cal A}$ is the anti\-symmetrization operator while 
the $\phi _p^k$ are the polarized pseudostates that describe the dipole, quadrupole, and octupole polarization
of the $3p^6$ core, respectively. Their structure and the corresponding polarizabilities are discussed in 
\cite{zatsarinny2010}.
The unknown functions $P(nl)$ for the outer valence electron
were expanded in a \emph{B}-spline basis, and the corresponding
equations were solved subject to the condition that the wave
function vanishes at the $R$-matrix boundary, which is chosen such that 
exchange effects between the projectile and the target electrons outside the box are negligible. 
The \emph{B}-spline coefficients for
the valence orbitals $P(nl)$ were obtained by diagonalizing the
\emph{N}-electron atomic Hamiltonian. 
We included 165 \emph{B}-splines of order~8 in the present calculations.
Choosing $a=80\,a_0$ (with $a_0 = 0.529\times 10^{-10}\,$m denoting the Bohr radius),
we obtained a good description for all low-lying states of K up to $8s$ \citep{zatsarinny2010}
regarding both level energies and oscillator strengths.  The deviations in the 
recommended excitation energies \citep{nist18} were 
less than approximately $0.01$ eV for all levels. Nevertheless, in the subsequent scattering calculations 
the calculated excitation energies were adjusted to the experimental ones 
to remove any uncertainties related to the slightly different excitation thresholds.

The above scheme is also able to generate continuum pseudostates
that lie above the ionization threshold. 
The scattering calculations were carried out by using a fully
parallelized version of the BSR complex \citep{zatsarinny2006}. 
These $R$-matrices with pseudostate calculations are similar to the recent BSR calculations
for e-Be \citep{zatsarinny2016} and e-Mg \citep{barklem2017} collisions.  More computational details are given in those papers.
The final close-coupling expansions contained 284 target states, including 30 bound states 
plus 254 continuum pseudostates with orbital angular momenta up to $L=4$, which covered the 
target continuum up to 50 eV above 
the first ionization limit.
We calculated partial waves for total orbital angular
momenta up to $L_{max} = 50$ numerically. Overall, with the various
total spins and parities, this involved 204 partial waves.
We considered all transitions between the lowest 17 physical states.  The principal difficulty 
for initially excited states was the
slow convergence of the partial wave expansions for transitions
between close-lying levels. When needed, we employed a top-up procedure
based on the Coulomb-Bethe approximation.

\subsection{e+K rate coefficients}
\label{ek_data}

In non-LTE applications, the rate coefficient is required, which is calculated by folding the cross sections $\sigma$ 
produced in the CCC and BSR calculations with the velocity distribution, assumed here to be the Maxwell 
distribution. The relevant equations are given in \cite{barklem2017}, and the data presented here are similar in 
form. The effective collision strengths $\Upsilon_{ij}$ from the CCC and BSR methods are calculated for 
transitions between the 15 lowest-lying states of K, which includes all states up to $6d$ at $3.93$ eV; this is 
the complete set of low-lying states that are included in both calculations.  The states and their experimental energies are listed in Table \ref{tab:energies}. The effective collision strength calculations are done 
for temperatures $T$ ranging from $1000$ K to $10000$ K in steps of $1000$ K, with additional results for 
$500$ K at the cool end, and $15000$ K and $20000$ K at the hot end. The data are only provided 
electronically at CDS. The data presented there are given as matrices, following the ordering of indexes in 
Table \ref{tab:energies} (i.e., the transition 1--2 corresponds to element (1,2)), one matrix for each temperature.   

The data from the CCC and BSR methods are generally in very good agreement, with the location (offset) and 
scale (scatter) of the ratio $\Upsilon_{ij}(\mathrm{CCC})/\Upsilon_{ij}(\mathrm{BSR})$, assuming a 
log-normal distribution, 1.03 and 0.17, respectively (see \cite{barklem2017}).  This indicates a mean offset of only 
3\% and scatter of 17\%.  In Figure \ref{rate_coeff_ek}, we compare the CCC and BSR data to each other and to older calculations.

\begin{table}
\begin{center}
\caption{Physical states for which data are provided from the CCC and BSR calculations, 
along with experimental excitation energies ($E_\mathrm{expt}$) from \cite{nist18}.}
\label{tab:energies}
\begin{tabular}{rrrr}
\hline \hline
Index & State & g & $E_\mathrm{expt}$ \\
      &       &   &  [eV]             \\
  1 &         $4s$ &   2 &  0.000   \\
  2 &         $4p$ &   6 &  1.615   \\
  3 &         $5s$ &   2 &  2.607   \\
  4 &         $3d$ &  10 &  2.670   \\
  5 &         $5p$ &   6 &  3.064   \\
  6 &         $4d$ &  10 &  3.397   \\
  7 &         $6s$ &   2 &  3.403   \\
  8 &         $4f$ &  14 &  3.487   \\
  9 &         $6p$ &   6 &  3.596   \\
 10 &         $5d$ &  10 &  3.743   \\
 11 &         $7s$ &   2 &  3.754   \\
 12 &         $5f$ &  14 &  3.795   \\
 13 &         $5g$ &  18 &  3.796   \\
 14 &         $7p$ &   6 &  3.853   \\
 15 &         $6d$ &  10 &  3.930   \\
\hline \hline 
\end{tabular}
\end{center}
\end{table}

\section{Non-LTE model}
\label{atom}

We performed non-LTE modeling in the trace-element approximation of the optical spectral lines of \ion{K}{I} 
using 1D plane-parallel MARCS \citep{gustafsson2008} model atmospheres. 
The statistical equilibrium and spectra were calculated using  version 2.3 of the radiative transfer 
code MULTI \citep{carlsson1986,carlsson1992}.

\subsection{Energy levels and radiative transitions}

 \begin{figure}[!htp]
                \centering
        \includegraphics[width=9.09cm]{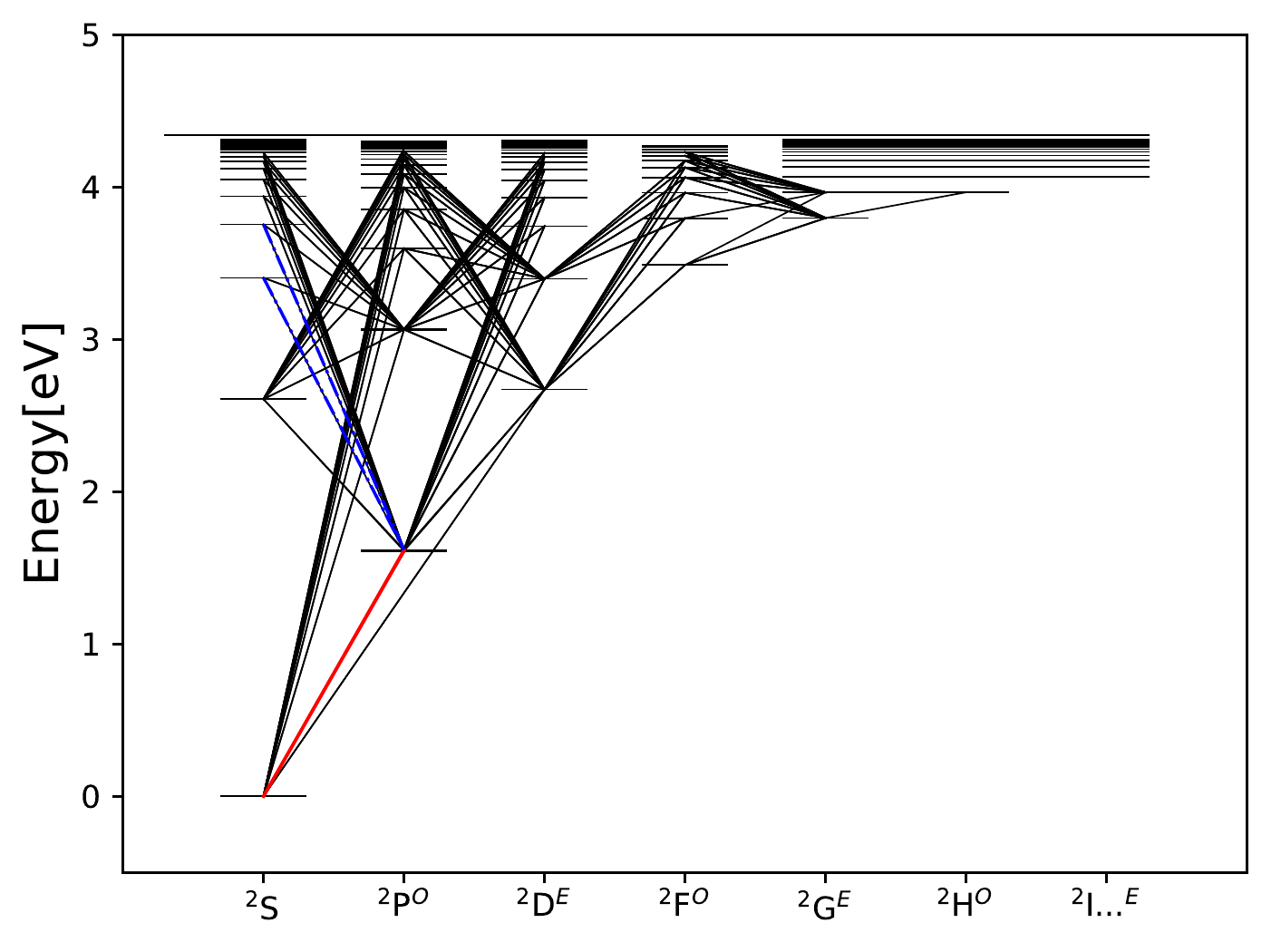}
    \caption{Grotrian diagram of \ion{K}{I}. In solid red we indicate 
    resonance transitions (both $7664 \ \AA$ and $7698 \ \AA$) and 
    the blue dashed lines the $5801 \ \AA$ and $6939 \ \AA$ transitions.}
    \label{grotion}
\end{figure}

Our potassium model is complete up to $0.13$ eV below the first ionization energy ($4.34$ eV), 
with all available levels with configurations up to principal quantum number $n=20$ and the \ion{K}{II} ground level. 
We have a total of $134$ levels in our atom, out of which $110$ are fine-structure-resolved energy 
levels from the \cite{nist18} database, which comes primarily from \citet{sugar1985} and \citet{sansonetti2008}. 
The \cite{nist18} database is complete up to orbital quantum number $l=3$, and the missing high-$l$ data 
were computed under the assumption that they are Rydberg levels ($24$ nonfine-structure-resolved levels).

The oscillator strengths of the allowed bound-bound radiative transitions were taken from
\cite{wiese1969}, \cite{biemont1973}, and \cite{sansonetti2008}. The total number of bound-bound 
transitions considered is 250. The Grotrian diagram of our atomic model can be seen in Figure \ref{grotion}. 
All bound-bound transitions involving the Rydberg levels are disregarded as it was seen that transitions with 
wavelengths larger than around $20000 \ \AA$ had negligible impact on the statistical equilibrium. 
Therefore, these levels were added so that the code computes the partition function as accurately as possible.
Whenever available, broadening data were added from the VALD database \citep{vald2015} or  
from \cite{barklem1998a}. 

\subsection{Photoionization cross sections}

The photoionization cross-sections of all levels between 4s and 7s are fine-structure resolved and were taken 
from \cite{zatsarinny2008} and \cite{zatsarinny2010}, calculated using the fully relativistic DBSR method. The 
photoionization 
cross-sections of the remaining levels were calculated using the hydrogenic approximation 
for bound-free transitions \citep[][ Eq. 8.4]{gray2005}. For higher levels, the cross sections of the two methods 
are more compatible with each other compared to the lower levels, and they start to diverge for 
increasingly higher wavelengths. Details and examples 
of the cross-sections can be seen in Figures 1-6 of \cite{zatsarinny2010}.


\subsection{Collisional data}
\label{coll_data}

\begin{figure}[!htp]
                \centering
        \includegraphics[width=9.09cm]{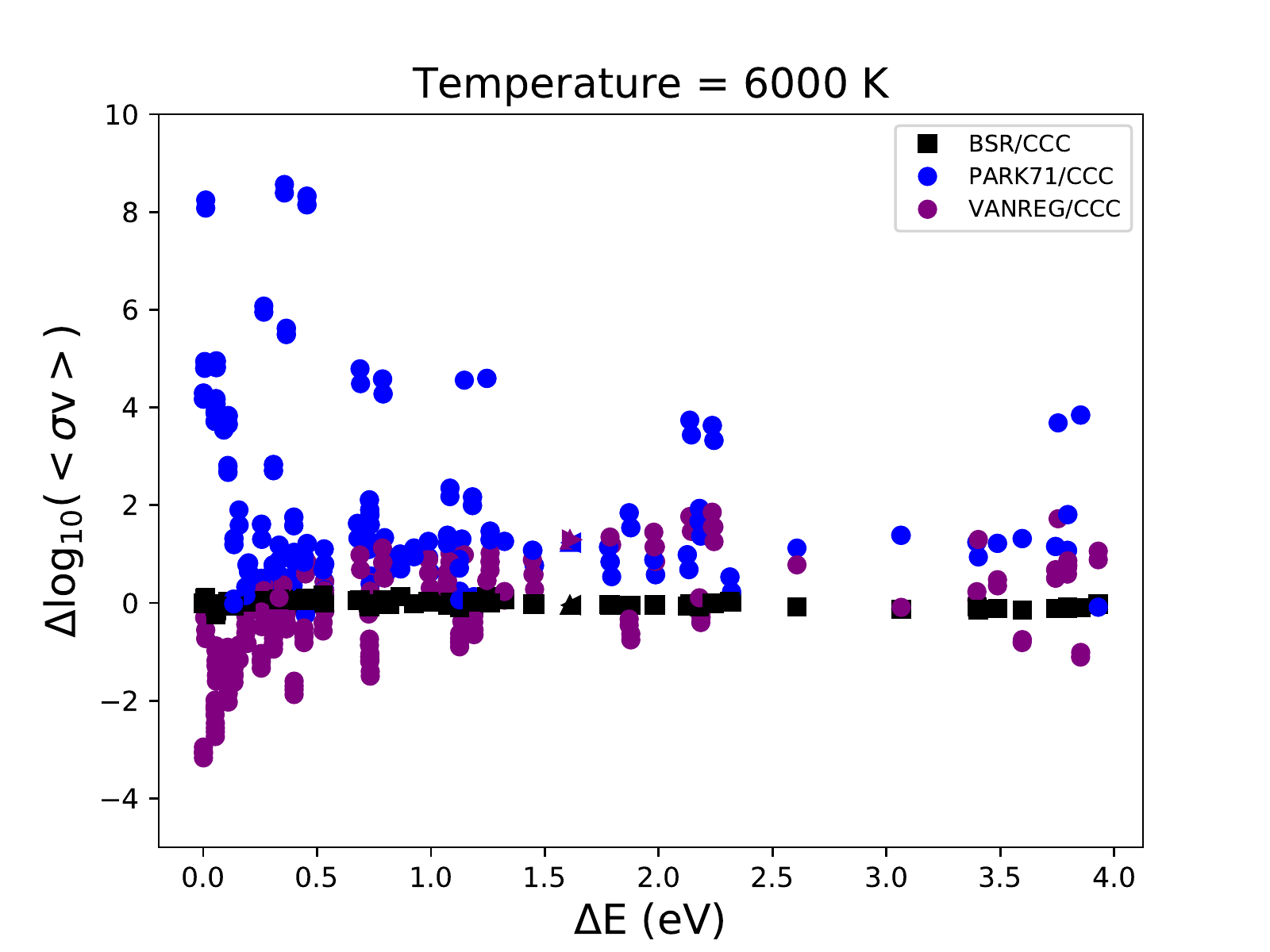}
    \caption{Ratio between e+K rate coefficients calculated through different methods. The triangles are the 
    rates of the resonance transitions.}
    \label{rate_coeff_ek}
\end{figure}

The inelastic e+K collisions have a large  impact on the statistical equilibrium of potassium. 
As discussed in Sect. \ref{ecol}, the inelastic e+K collision rates for transitions between low-lying levels 
4s and 6d were calculated using the CCC and BSR methods. Although CCC was employed in our 
standard atom, the e+K rates calculated via BSR were also tested, and the results with either 
method are indistinguishable; see Sect. \ref{nlte_effects_tests}. The rate coefficients for transitions from 
or to levels higher than 6d were all calculated using the Park71 method.

As mentioned in Sect. \ref{ek_data}, in Figure \ref{rate_coeff_ek} we compare the e+K rate coefficients 
calculated with the CCC,  BSR, Park71, and vanReg62 methods at $6000$ K.
We can see that CCC and BSR are in very good agreement, while the Park71 and vanReg62 methods differ from the two more recent and accurate methods. These disagreements lead to 
differences in the statistical equilibrium, and hence the synthetic lines, which are further 
explored in Sect. \ref{nlte_effects_tests}. We note that the rate coefficients of the resonance transitions  
4p$^{3/2}$ - 4s$^{1/2}$ and  4p$^{1/2}$ - 4s$^{1/2}$  calculated via both the Park71 and the 
vanReg62 methods are very similar to each other (blue and purple triangles in Fig. \ref{rate_coeff_ek}).

The inelastic H+K collisions also influence the 
statistical equilibrium of potassium.  For transitions involving low-lying levels (up to level 4f), we 
adopted the rate coefficients from \cite{yakovleva2018}, which are based on the LCAO 
model of \cite{barklem2016}. Following \cite{amarsi2018}, we added these data to rate 
coefficients calculated using the free electron model in the scattering length 
approximation \citep[Eq. 18,][]{kaulakys1991}. The rate coefficients for transitions from or 
to levels higher than 6d were all calculated using the free electron model alone.


Finally, inelastic e+K collisional ionization was calculated using the empirical formula in chapter 3 of 
\cite{allen1976}. These rates are important for guaranteeing LTE populations deep in the stellar atmosphere.

\section{Non-LTE effects on Potassium}
\label{nlte_in_potassium}

\subsection{Test models}
\label{nlte_effects_tests}

\begin{table*}
  \centering
  \caption[]{Different atoms created to compare the collisions and photoionization recipes.}
  \label{atoms}
  \tiny
  \begin{tabular}{llllllllllllllllllllllll}
     \hline
     \hline
     \noalign{\smallskip}
     Atom designation & & & & & & & Electron excitation  & & & & & Neutral hydrogen collisions  & & & & & & & photoionization & & & & \\
\end{tabular}\\
\begin{tabular}{lccccccccc}
      & CCC & BSR & Park71 & vanReg62 & \cite{barklem2016} & \cite{kaulakys1991} & \cite{zatsarinny2010} & 
      Hydrogenic  \\
     \noalign{\smallskip}
     \hline
     \noalign{\smallskip}
        \emph{a})Standard atom & X & & X & & X & X & X & X \\
        \emph{b})BSR e+K & & X & X & & X & X & X & X \\
        \emph{c})Park71 e+K & & & X & & X & X & X & X \\
        \emph{d})Hydrogenic photoionization & X & & X & & X & X & & X \\
        \emph{e})Resonance transitions only & X & & X & & X & X & X & X \\       
        \emph{f})No resonance transitions & X & & X & & X & X & X & X \\ 
        \emph{g})Reduced H+K excitation & X & & X & & x$10^{-3}$ & x$10^{-3}$ & X & X \\
        \emph{h})Reduced e+K excitation  & x$10^{-3}$ & & x$10^{-3}$ & & X & X & X & X \\
        \emph{i})vanReg62 e+K & & & & X & X & X & X & X \\
    \noalign{\smallskip}
     \hline     
  \end{tabular}
\end{table*}

Below we present the results of test calculations for the cases of different inelastic e+K collisional rates 
(CCC, BSR, Park71, and the vanReg62 calculations), and
different photoionization calculations (the DBSR method by \cite{zatsarinny2010} and the hydrogenic 
approximation). Our standard atom is atom \emph{a}, discussed in Sect.~\ref{atom}. The collisions employed 
in each 
atom can be seen in Table \ref{atoms}. Atoms \emph{e} and \emph{f} have the same collisions as the standard 
atom but 
the first only has the resonance 4p$^{3/2}$ - 4s$^{1/2}$ and  4p$^{1/2}$ - 4s$^{1/2}$ radiative 
transitions and the second has all other transitions except those.
Atom \emph{g} has the H+K excitation rates decreased by a factor of $10^{-3}$; 
and atom \emph{h} has the e+K excitation rates decreased by a factor of $10^{-3}$. The last atom (atom 
\emph{i}) uses the vanReg62 e+K collisions.

\subsection{Departure coefficients}
\label{nlte_effects_departures}

\begin{figure}[!htp]
                \centering
        \includegraphics[width=9.09cm]{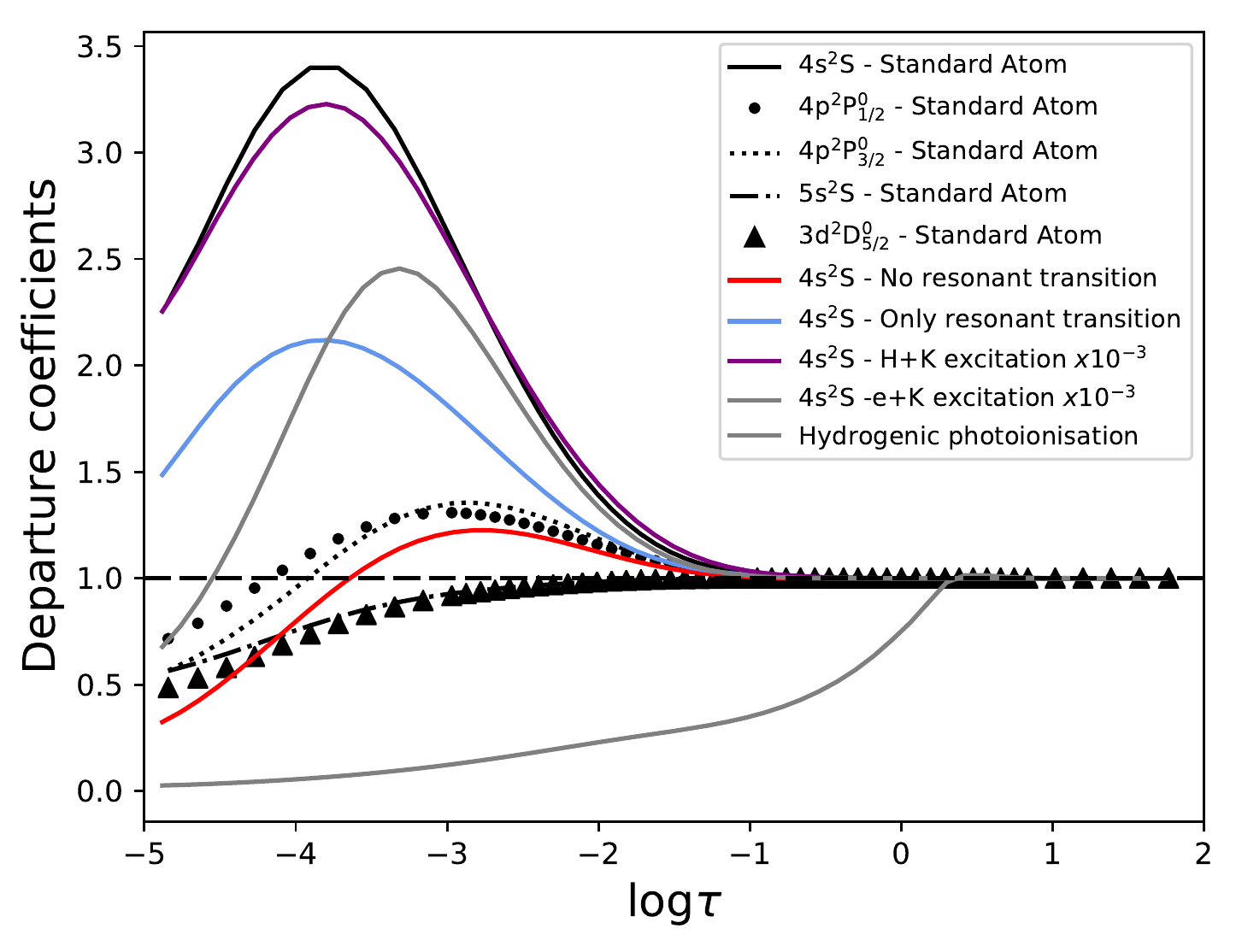}
    \caption{Departure coefficients for the low-excited \ion{K}{I} levels in the solar atmosphere.}
    \label{dptcoef}
\end{figure}

In Figure \ref{dptcoef} we show a number of departure coefficients, 
b$_{\rm{k}}$ = n$_{\rm{NLTE}}$ /n$_{\rm{LTE}}$, in the solar atmosphere, for our standard atom and 
for a number of our test atoms (Sect.~\ref{nlte_effects_tests}), designed to 
test the main contributors to the non-LTE effects in potassium.

From Figure \ref{dptcoef} and from previous works \citep[e.g.,][]{zhang2006}, we see there is a 
strong overpopulation of the \ion{K}{I} ground state and the two first excited states (4s, 4p$^{1/2}$ and 
4p$^{3/2}$). 
The major drivers of this overpopulation are the resonance transitions due to photon losses. 
This was previously demonstrated in Sect. 4.3 of \citet{bruls1992}. According to their study, 
the infrared lines are also important in the statistical equilibrium (but to a much lower degree) and 
as such, Figure \ref{dptcoef} illustrates that there are departures from LTE even when the resonance 
transitions are switched off (test atom \emph{f}).

The departure coefficients also illustrate that the inelastic H+K excitation is of lesser importance on 
solar metallicity than the inelastic e+K excitation, in line with previous studies for alkali metals 
\citep{lind2009,lind2011}. This is further discussed in the following section through the analysis of
synthetic lines.

\subsection{Effects on spectral lines}

\begin{figure}[!htp]
                \centering
        \includegraphics[scale=0.77]{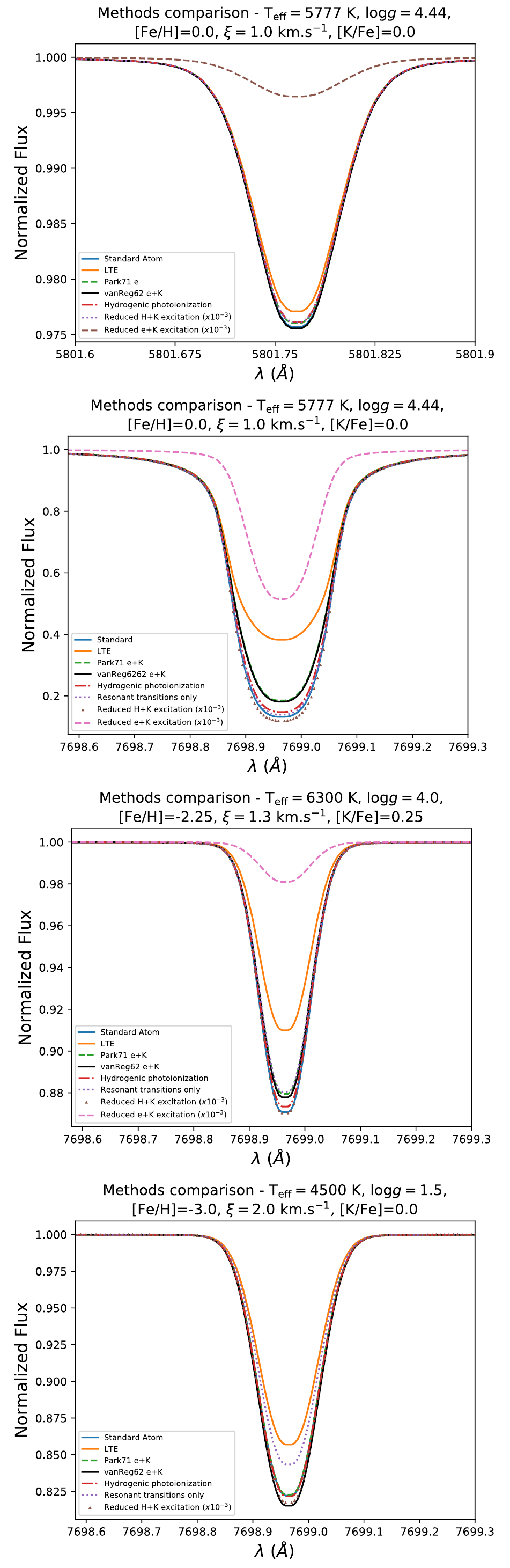}
   \caption{Comparison between the synthetic spectral lines using 
   different collisional recipes for the $5801 \ \AA$ and $7698 \ \AA$ lines in the 1D solar atmosphere (first and second panels). 
   The third and fourth panels show the same but for a 1D metal-poor atmosphere (HD 84937) and a giant star 1D atmosphere, respectively.}
  \label{line_comp}
\end{figure}

To generate the synthetic spectral lines we employed MARCS \citep{gustafsson2008} 1D model atmospheres 
of the Sun with a potassium abundance of $A\rm{(K)}=5.11$, a metal-poor 
star with the same stellar parameters determined for HD 84937 in \cite{peterson2017}, and 
adopted by \cite{spite2017}, with 
a potassium abundance of $A\rm{(K)}=3.08$ ($\teff = 6300$ K, log$g=4.00$ , [Fe/H]$=-2.25$ and 
$\xi=1.3 \ \rm{km.s^{-1}}$), and a giant star with $\teff = 4500$ K, log$g=1.5$, 
[Fe/H]$=-3.00$, $\xi=2.0 \ \rm{km.s^{-1}}$, and [K/Fe]$=0.0$.

We first discuss how the different collisional and radiative data affect line formation in the solar atmosphere. 
The results for the $5801 \ \AA$ and $7968 \ \AA$ lines are shown in the first and second panels of 
Figure \ref{line_comp}. We note that for all tests described in this section, the standard atom (\emph{a}) 
and the atom replacing CCC collisions with BSR collisions (\emph{b}) 
produce indistinguishable line profiles; as such we do not plot the results of the latter atom.

In our analysis, the high-excitation $5801 \ \AA$ line is almost insensitive to departures from LTE. Consequently 
it is insensitive to the details of the non-LTE modeling, with variations of only around $0.01$ dex. 
Unfortunately, the line is very weak and can only be detected in very high-resolution and high-S/N 
spectra of stars with solar metallicity or higher, and solar effective temperature or cooler. Therefore, it is 
unsuitable for abundance analysis in representative samples. This  $5801 \ \AA$ line is however 
very sensitive to the large decrease in e+K excitation of atom \emph{h}, a case in which the line is almost not 
formed and is much weaker than in LTE.

The resonance line has a different behavior. While the wings are insensitive to departures from LTE, 
the core is significantly deeper in non-LTE, as a result of the overpopulation of the ground state 
as we discussed in Sect. \ref{nlte_effects_departures}. There is a non-negligible 
difference between the line cores when using different collisional recipes. 

Concerning the sensitivity of the non-LTE effects to the atomic data, the most prominent difference to the 
resonance line is observed when we decrease the e+K excitation rates by $10^{-3}$ (atom \emph{h}). 
In this case the statistical equilibrium changes to the point where the line is not as deep as the LTE case, due to 
an increased importance of the photoionization -- 
an effect also observed in the departure coefficient. 
The second largest difference is seen when the CCC collisions are replaced with the Park71 or vanReg62 
collisions (atoms \emph{c} and \emph{i}): 
an abundance difference of $\Delta = +0.11$ dex with respect to the standard atom (\emph{a}). 
The third largest difference is seen when adopting 
hydrogenic photoionization cross-sections (\emph{d}): an abundance difference of 
$\Delta = +0.03$ dex with respect to the standard atom (\emph{a}). 
When synthesizing the line with inefficient H+K collisions (\emph{g}) the line core is deeper than the line 
from the standard atom (\emph{a}), but the 
absolute abundance difference is the same as that produced by the atom with hydrogenic photoionization (atom 
\emph{d}). The atom with only two radiative transitions, namely the resonance lines (\emph{e}), has 
virtually the same abundance as our standard atom. 

The non-LTE effects are mainly a source-function effect caused by photon losses (resonance scattering). 
Like in the Na D lines \citep[see Sect 3.1 of][]{lind2011}, the overpopulation of the ground state pushes the mean 
formation depth outward, deepening the lines -- but only slightly. The main effect that deepens the spectral line 
is the sub-thermal line source function. The line source function of a line formed by pure resonance scattering 
is determined by the radiation field, and therefore by the radiative rates in the lines themselves. This is why 
the atom with only the resonance lines (\emph{e}) performs so similarly to the standard atom (\emph{a}), 
that is, because the source function is the same in the two cases.   

In the third panel of Figure \ref{line_comp} we show the synthetic spectral lines in 
a metal-poor turn-off star. Similarly to the solar atmosphere case, for HD 84937, compared to the standard 
atom (\emph{a}), the non-LTE effects change the most when the e+K collisions are reduced (\emph{h}), 
whereas reducing the H+K collisions (\emph{g}) has only a small impact.
The largest abundance differences ($\Delta = +0.04$ dex) are for atoms \emph{c}, \emph{e}, 
and \emph{i}. The third largest is for atom \emph{d} ($\Delta = +0.01$ dex).
The general difference between the synthetic lines in HD 84937 is smaller than in the solar 
atmosphere, because in a metal-poor atmosphere the hydrogen collisions are more relevant to the 
statistical equilibrium, owing to the reduced number of free electrons.

The stars analyzed in this work are late-type stars, in which the H+K excitation is expected to be less 
important than in giant stars. In the last panel of Figure \ref{line_comp} we tested the differences in a giant star. 
In this case the lines using the standard and the \emph{c} atoms are indistinguishable. This happens because 
in this metal-poor giant star there are not as many free electrons and the contribution of this process 
becomes less important. Although less important, the effect of the vanReg62 e+K collision 
rates, which are mostly higher than the other methods employed, can be observed as it increases the depth of 
the line core. Nevertheless, as the importance of the e+K collisions decreases, we observe that the importance 
of the H+K excitation increases: compared to the 
standard atom (\emph{a}), there is an abundance difference of $\delta\approx0.03\,\dex$ after reducing the 
efficiency of the H+K collisions (\emph{g}). This can be contrasted with the late-type metal-poor star HD 84937 
(third panel of Figure \ref{line_comp}), where no appreciable difference is observed.  

In this section, we showed that the use of improved calculations of electron collisions is the most influential factor 
in the non-LTE line profile of dwarfs. The CCC and BSR calculations give comparable results, 
but substituting these calculations with the older Park71 and vanReg62 methods considerably changes the 
strength of the synthetic lines. 
Improved photoionization cross-section calculations with the DBSR method 
also had non-negligible differences in the final line depth of the synthetic lines, 
and are important for accurate abundance determinations through the $7698 \ \AA$ resonance line.

\section{Abundance analysis of K in benchmark stars}
\label{analysis}

\begin{table*}
  \centering
  \caption[]{Stellar parameters of the sample.}
  \label{stellar_info}
  \small
  \begin{tabular}{lrrrrc}
     \hline
     \hline
     \noalign{\smallskip}
     Star & T$_{\rm{eff}}$ (K) & $\log \ g$ ($g$ in $\rm{cm.s^{-2}}$) & $\xi$ ($\mathrm{km\,s^{-1}}$) & [Fe/H] & Broadening* ($\mathrm{km\,s^{-1}}$)
     \\
     \noalign{\smallskip}
     \hline
     \noalign{\smallskip}
        Sun & $ 5772^a $ & $ 4.44^a $ & $ 1.0^b $ & $ 0.00^c $ & $3.5$\\
        HD 192263 & $ 4998^d $ & $ 4.61^d $ & $ 0.66^e $ & $ -0.05^e $  & $3.0$\\
        Procyon & $ 6556^f $ & $ 4.01^f $ & $ 1.85^g $ & $ -0.02^h $  & $6.0$\\
        HD 103095 & $ 5140^i $ & $ 4.69^j $ & $ 0.9^k $ & $ -1.13^l $  & $1.0$\\
        HD 140283 & $ 5787^i $ & $ 3.66^m $ & $ 1.6^g $ & $ -2.28^n $ & $3.8$\\
        HD 84937  & $ 6371^d $ & $ 4.05^o $ & $ 1.3^p $ & $ -1.97^n $  & $5.0$\\
    \noalign{\smallskip}
     \hline     
  \end{tabular}
  \tablefoot{a) Reference value from \cite{prsa2016}; b) \cite{spina2016}; c) \cite{asplund2009}; 
  d) IRFM value from \cite{casagrande2011}; e) \cite{andreasen2017}; f) Fundamental value from 
  \cite{chiavassa2012}; g) \cite{pancino2017}; h) $<\rm{3D}>$ non-LTE \ion{Fe}{II} from \cite{bergemann2012}; 
   i) Fundamental value from \cite{karovicova2018}; j) Fundamental value from \cite{bergemann2008}; 
  k) \cite{reggiani2018}; l) 1D LTE \ion{Fe}{II} from 
  \cite{ramirez2013} with $<3D>$ non-LTE corrections from \cite{amarsi2016b}; m) \cite{gaiadr2}; 
  n) 3D non-LTE value from \cite{amarsi2016b}; o) Fundamental value from \cite{vandenberg2014}; p) 
  \cite{spite2017};* Includes both macroturbulence velocity and rotation in one Gaussian broadening 
  kernel.}
\end{table*}

\begin{table}
  \centering
  \caption[]{Atomic data of the transitions}
  \label{transitions_info}
  \tiny
  \begin{tabular}{lrrrrr}
     \hline
     \hline
     \noalign{\smallskip}
     $\lambda \ (\AA)$ & Transition & E (eV) & log($gf$) & $\sigma / a^2_0$ & $\alpha$ \\
     \\
     \noalign{\smallskip}
     \hline
     \noalign{\smallskip}
                $5801.75$ & 4p$^2$P$^0_{3/2}$ - 7s$^2$S$_{1/2}$ & $1.617 - 3.753 $ & $-1.605$ & -- & --\\
                $6938.77$ & 4p$^2$P$^0_{3/2}$ - 6s$^2$S$_{1/2}$ & $1.617 - 3.403 $ & $-1.252$ & $1721$ & 
                $0.294$\\
                $7698.96$ & 4s$^2$S$_{1/2}$ - 4p$^2$P$^0_{1/2}$ & $0.000 - 1.610 $ & $-0.176$ & $485$ & 
                $0.232$\\
                $12522.16$ & 4p$^2$P$^0_{3/2}$ - 5s$^2$S$_{1/2}$ & $1.617         -       2.607$ & $-0.134$ & $1260$ & 
                $0.183$ \\
    \noalign{\smallskip}
     \hline     
  \end{tabular}
  \tablefoot{The broadening of the lines via elastic collisions with hydrogen are represented via 
  $\sigma$, the cross-section at the velocity of $10^4 \ \rm{m.s^{-1}}$, and $\alpha$, the exponent 
  with which the cross-section varies with velocity \citep[$v^{-\alpha}$, ][]{anstee1995}, and both $\sigma$ and 
  $\alpha$ are from \cite{barklem1998a}. For the $5801 \ \AA$ we use the Unsold's method 
  \citep{unsold1955}, scaled to a factor of 1.5.}
\end{table}

We further tested our standard atom by modeling the \ion{K}{I} optical lines in different stellar atmospheres. 
We analyzed the Sun and the following benchmark stars of the GAIA-ESO spectroscopic survey: HD 84937, 
HD 103095, HD 192263, HD 140283 and Procyon. For the Sun we use the \cite{stenflo2015} flux 
solar atlas, and the spectra of the remaining objects are high-resolution (R$\approx 220000$) PEPSI spectra 
\citep{strassmeier2018}; these are available fully reduced and continuum normalized 
\footnote{\url{https://pepsi.aip.de/?page\_id=552}}. The stellar parameters of the stellar model 
atmospheres can be seen in Table \ref{stellar_info}.

To determine the potassium abundances of the stars we match synthetic equivalent widths (EWs) to observed ones. 
We tested both Gaussian fitting and full line integration for measuring the EWs in the observed 
spectra and the methods have similar outcomes. We convolved (Gaussian Kernel) the synthetic spectra  of the best 
abundance to account for rotational and macroturbulence velocities using the 
PyAstronomy\footnote{www.hs.uni-hamburg.de/DE/Ins/Per/Czesla/PyA/PyA/index.html} 
python package. Both are treated as a free parameter but for a first guess we calculate the macoturbulent velocity 
using the trend with $\teff$ described in \cite{gray2005}.

We analyzed three observable lines in the optical spectral region: the \ion{K}{I} $5801.7$, $6938.7,$ and 
$7698.9 \ \AA$ lines, and for completeness added the $12522 \ \AA$ infrared line in the Sun. The 
atomic data of the transitions are in Table \ref{transitions_info}.
We used standard MARCS \citep{gustafsson2008} model atmospheres. 
In contrast with \cite{zhang2006} and \cite{scott2015} we did not analyze the 
\ion{K}{I} $4044.1$ and $7664 \ \AA$ lines because they are heavily blended. 
We show the fitted lines in Figures \ref{figsun} to \ref{hd19}.

We estimate a lower bound on the modeling errors through the line-to-line scatter of the three observable lines 
in the Sun and HD 192263. In the Sun, the scatter is only $0.02$ dex, but it is up to $0.05$ dex 
in HD 192263, possibly due to the strong damping wings and difficulties in determining the 
abundance through the resonance line in this star (both are further discussed below). Most of our analyzed 
stars do not show such strong damping wings and asymmetries observed in HD 192263, and therefore the uncertainties 
are not as high. Thus, we estimate a lower error of $0.03$ dex for our measurements. 

\subsection{Sun}
\begin{figure*}
\centering
\includegraphics[width=17cm]{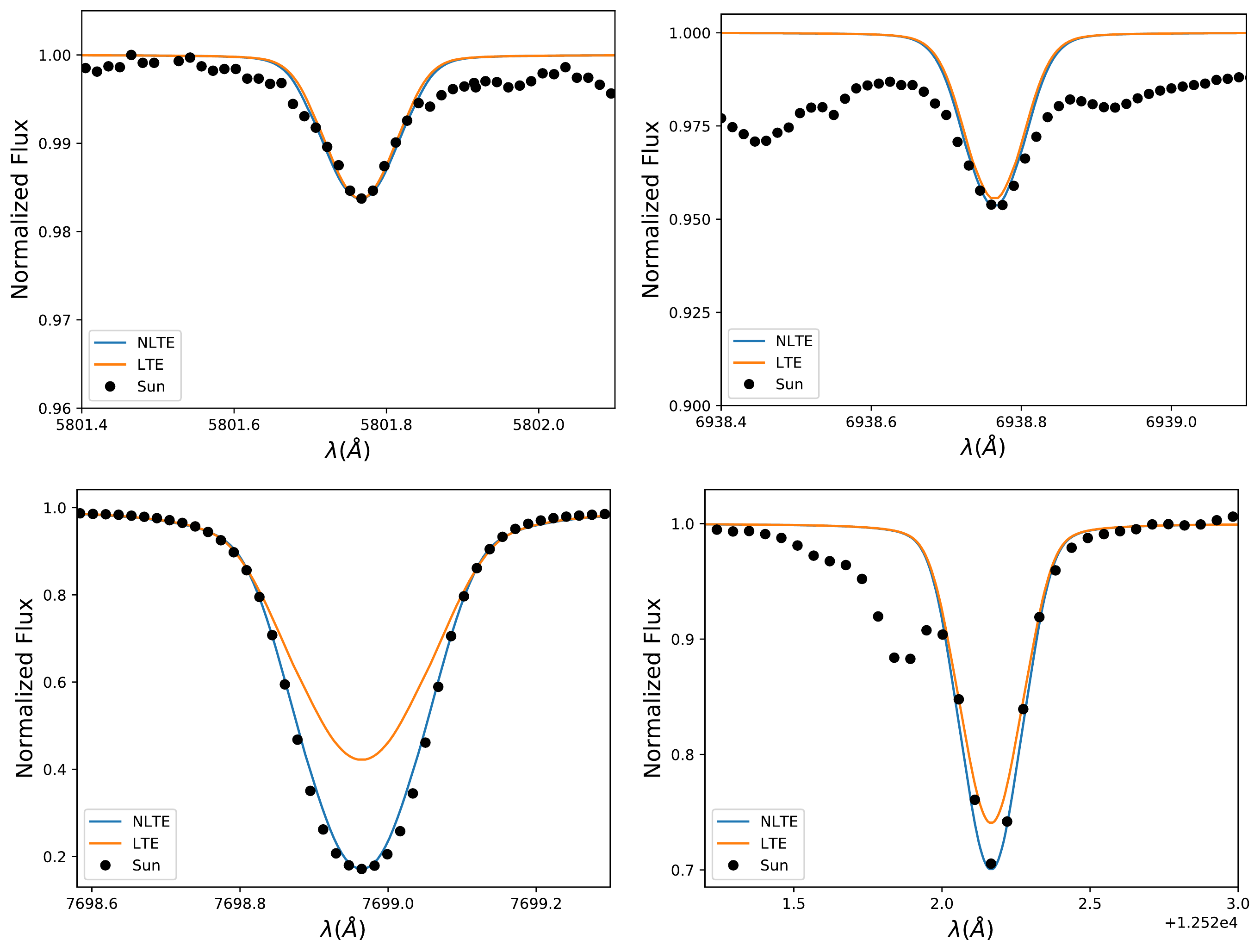}
\caption{Non-LTE abundance fit of the $5801$, $6939$, $7698$, and $12522 \ \AA$ lines (top left, top right, lower left, and lower right, respectively) at the Sun with an abundance of $A\rm{(K)}=5.11$. The LTE lines of the same abundance are also shown in the plots.}
\label{figsun}
\end{figure*}   

        For the Sun we derived a 1D non-LTE abundance of $A\rm{(K)}=5.11$ from averaging the result of the best 
        EW match of the three optical lines in the solar spectrum. In our analysis the individual abundances 
        of each line found were $A\rm{(K)}_{\rm{5801 \ \AA}} = 5.10$ dex, $A\rm{(K)}_{\rm{6939 \ \AA}} = 5.10$ 
        dex, and $A\rm{(K)}_{\rm{7698 \ \AA}} = 5.12$. 
        \cite{zhang2006} did not find, for these same lines, 
        abundances in such good agreement. Their abundances vary up to $0.09$ dex between the 
        optical ($A\rm{(K)}_{\rm{5801 \ \AA}} = 5.15$, 
        $A\rm{(K)}_{\rm{6939 \ \AA}} = 5.06,$ and $A\rm{(K)}_{\rm{7698 \ \AA}} = 5.14$). 
        With 1D MARCS model atmospheres (such as those used here) \cite{scott2015} 
        analyzed the lines $5801 \ \AA$ and $6939 \ \AA$, finding
        $A\rm{(K)}_{\rm{5801 \ \AA}} = 5.15$ with a non-LTE correction of $-0.03$ dex, in good agreement 
        with our abundance. For the other line in common they determined $A\rm{(K)}_{\rm{6939 \ \AA}} = 5.09$, 
        and applied a correction of $-0.03$ dex, a somewhat smaller abundance than what we found ($0.04$ dex) 
        but still within their expected error of $0.05$ dex for potassium.
        
        For our plot we needed to apply a change in the original normalization of the $5801$ 
        and $6939 \ \AA$ lines. As already pointed out by \cite{zhang2006} there are terrestrial blends 
        around the $5801 \ \AA$ line and a very uncertain continuum at the $6939 \ \AA$ region due to 
        a number of contributions from different lines. We optimized the continuum of those lines based on adjacent 
        regions ($5811 \ \AA$ and $6935 \ \AA$).
        
        It must be noted that adjacent features were not synthesized, and the potassium lines at 
        $5801$, $6939$, and $12522 \ \AA$ were synthesized with the best abundance 
        and a broadening parameter determined via the fitting of the $7698 \ \AA$ line. 
        As mentioned above, the abundances were not calculated via synthetic spectra. We separately measured the 
   EW of each observed line and matched them to the EWs of the synthetic 
   lines to find the abundances. Thus, the fits shown in Figure \ref{figsun} were not used to estimate 
   the abundances, and the line profiles are shown only to demonstrate that our EW abundance 
   can reproduce the lines.
        As can be seen, the synthetic line fits to the solar spectra are very good. We also show the LTE line of the 
        same abundance 
        as a comparison and one can see that although the differences between the LTE and non-LTE methods for the 
        $5801 \ \AA$ and $6939 \ \AA$ lines are very small (to fit the lines with LTE one needs an 
        abundance $\approx0.01$ dex 
        higher than the non-LTE abundance), the difference observed for the resonance $7698 \ \AA$ line is very 
        large. In the case of the resonance line, the LTE assumption fails completely and it is not possible to correctly 
        reproduce the spectral line, even when considerably increasing the abundance of \ion{K}{I} 
        to $A\rm{(K)}\approx 5.44$ (our best LTE abundance via EW). We emphasize the importance of taking non-LTE 
        effects into account when studying the GCE of potassium using the resonance line 
        \citep[e.g., ][]{takeda2002,zhang2006,takeda2009,andrievsky2010,scott2015}.

\subsection{HD 84937}

\begin{figure}[!htp]
                \centering
                \includegraphics[width=9.09cm]{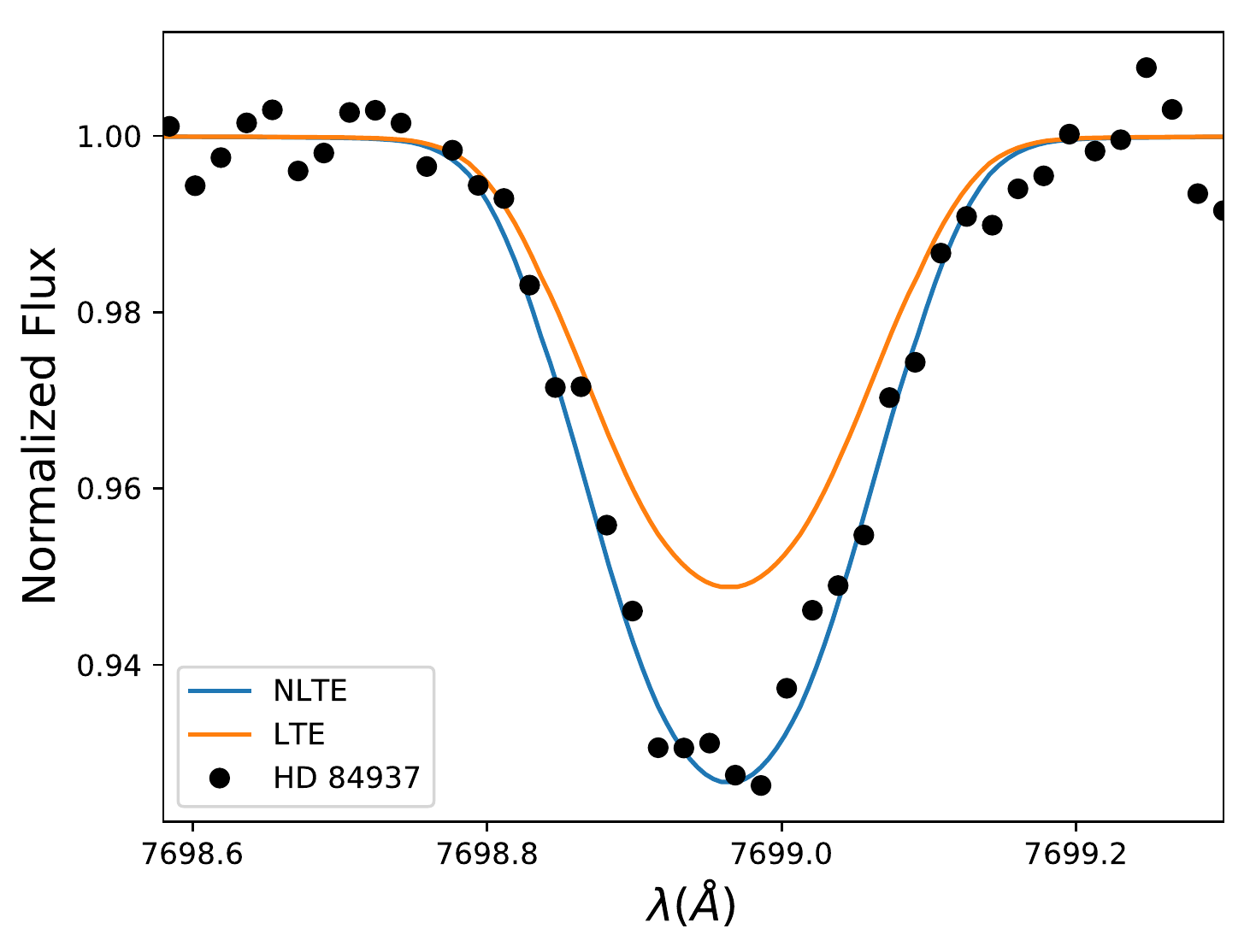}
            \caption{Non-LTE abundance fit of the $7698 \ \AA$ line for HD 84937 with an abundance of 
            $A\rm{(K)}=3.15$. The LTE line of the same abundance is also shown in the plot.}
            \label{hd082}
\end{figure}

HD 84937 is a low-metallicity ([Fe/H]$=-1.97$), bright, main sequence turn-off star that is 
commonly used as a standard representative of the abundance pattern of similar-metallicity field stars.
        
\cite{spite2017} analyzed the abundance pattern of this star and measured a LTE 
potassium abundance of $A\rm{(K)}=3.40$ from the resonance $7698 \ \AA$ line. These latter authors applied a non-LTE abundance correction of $-0.2$ dex from \cite{zhang2006}, finally advocating 
$A\rm{(K)}=3.20$ for this line in HD 84937.
        
In non-LTE, we measured $A\rm{(K)}=3.15$, while our best LTE abundance is $A\rm{(K)}=3.33$, 
a difference of $\Delta = -0.18$ dex, in excellent agreement with the correction from the model atom 
described in \cite{zhang2006} and used in \cite{spite2017}. In Figure \ref{hd082} we show the 
non-LTE abundance along with the LTE 
synthetic line of the same abundance for comparison. As can be seen, the non-LTE abundance that was found 
fits the stellar spectrum very well. The small asymmetry in the 
red wing of the observed spectral line is due to convective motions in the stellar atmosphere, 
the same conclusion as drawn from the analysis of this line by \cite{smith2001}. The best LTE line 
was also able to reproduce the observed line and the best non-LTE is indistinguishable from the 
best LTE line when plotted together.

\subsection{HD 103095}
\label{hd10sec}

\begin{figure}[!htp]
                \centering
                \includegraphics[width=9.09cm]{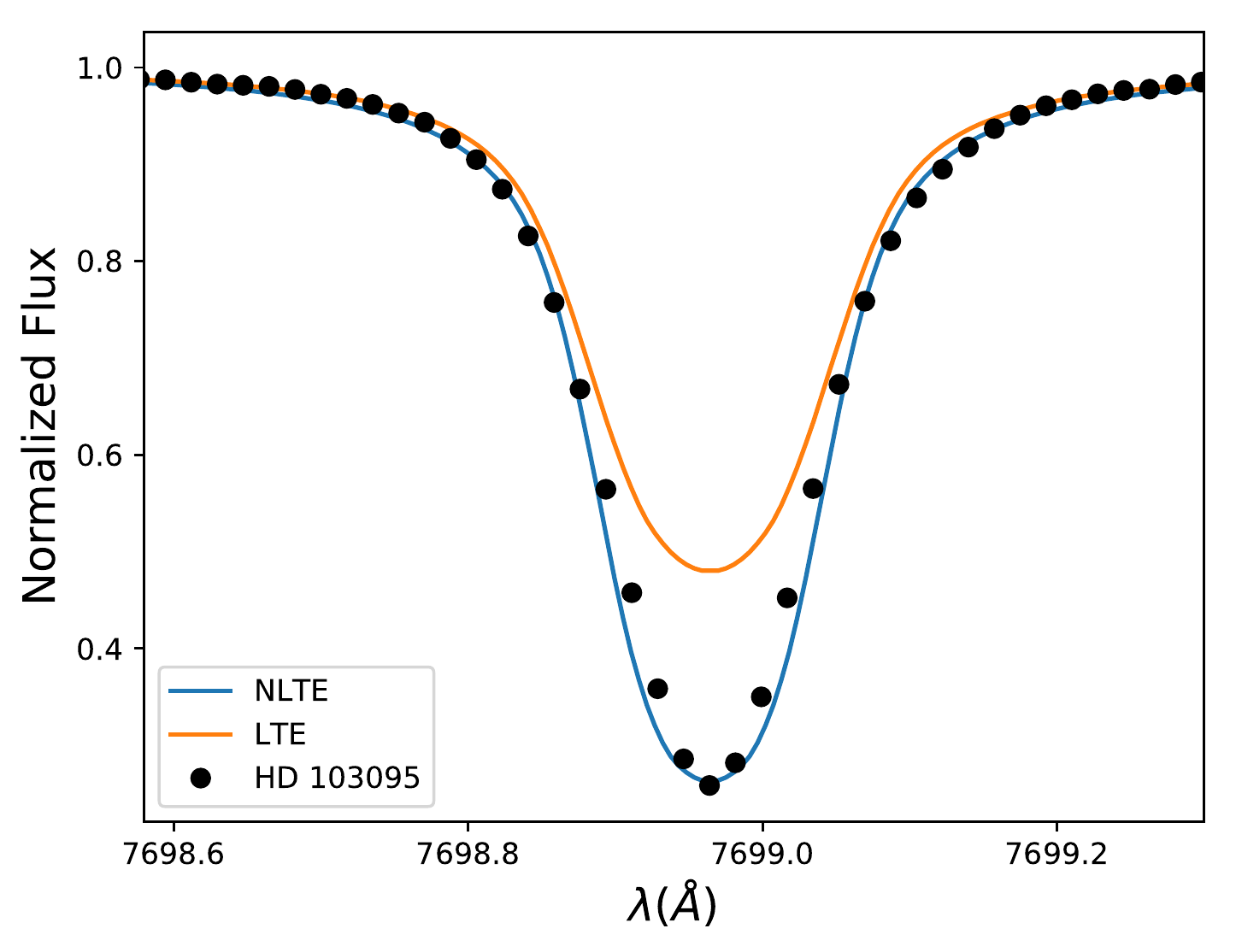}
            \caption{Non-LTE abundance fit of the $7698 \ \AA$ line for HD 103095 with an abundance of 
            $A\rm{(K)}=4.00$. The LTE line of the same abundance is also shown in the plot.}
            \label{hd10}
\end{figure}

        HD 103095 is a metal-poor K-type dwarf, commonly used as a standard star of the inner Halo. This star is 
        an $\alpha$-poor star with [Mg/Fe]$ = 0.12$ dex \citep{reggiani2018}, which is lower than usually found 
        for stars of such metallicity.
                
        When assuming LTE, we determined an abundance $A\rm{(K)}=4.26$ for this star.
        When the non-LTE modeling was used instead, the best match of EW was found for an abundance of 
        $A\rm{(K)}=4.00$, a difference of $-0.26$ dex. 
        The non-LTE synthetic line, the observed spectra, and the comparison LTE synthetic line can be seen in 
        Fig. \ref{hd10}. 
        The spectral line in HD 103095 is very well reproduced under non-LTE, except for 
        a small discrepancy in the near-to-core region, which might be caused by 3D effects (further discussed in 
        Sect. \ref{3dsec}).
        As for LTE, our best abundance estimate was not able to correctly reproduce the line.

\subsection{HD 140283}

\begin{figure}[!htp]
                \centering
                \includegraphics[width=9.09cm]{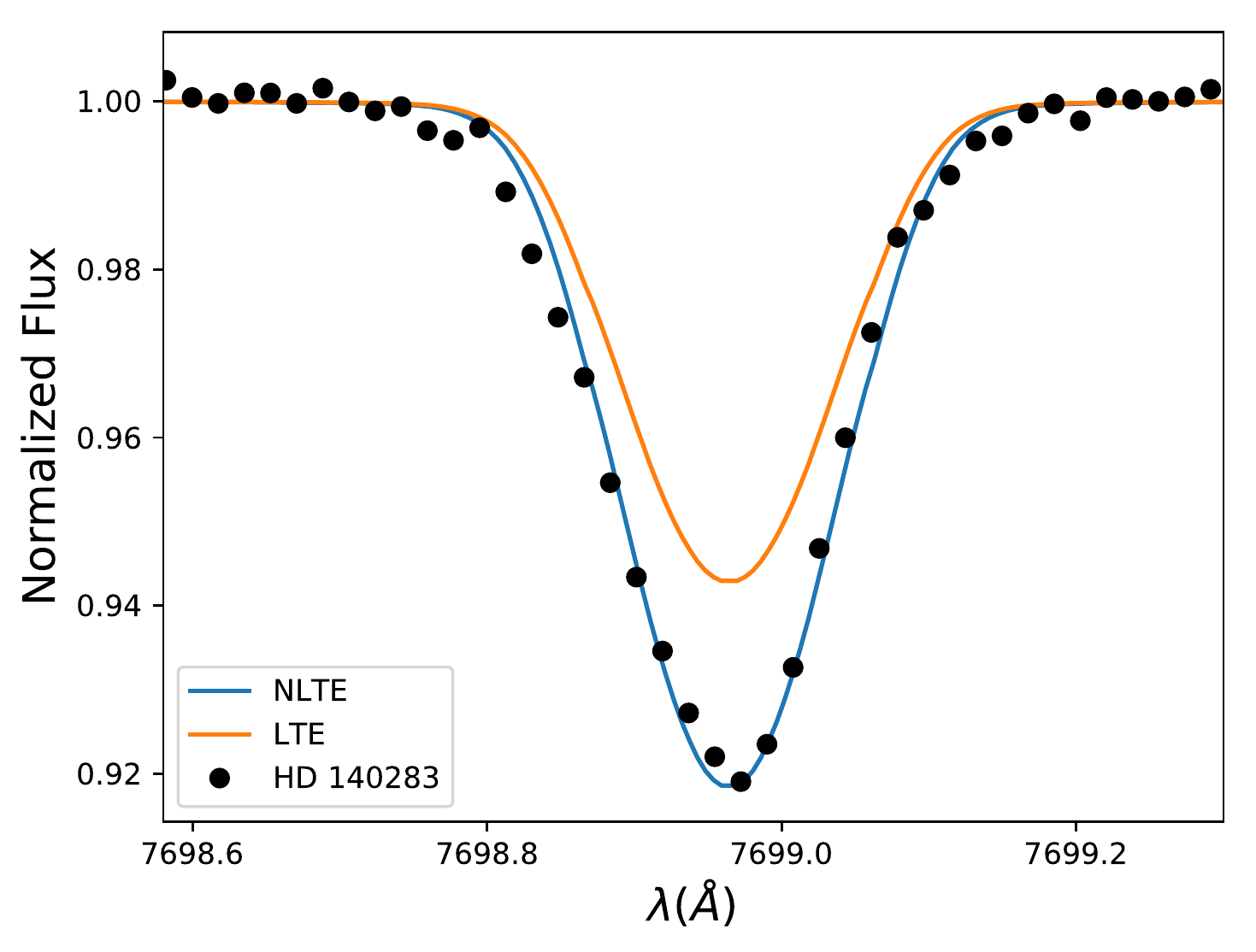}
            \caption{Non-LTE abundance fit of the $7698 \ \AA$ line for HD 140283 with an abundance of 
            $A\rm{(K)}=2.77$. The LTE line of the same abundance is also shown in the plot.}
            \label{hd14}
\end{figure}

        HD 140283 is a bright very metal-poor, high-velocity subgiant in the solar neighborhood. Its proximity 
        to the Sun made it the first star with spectroscopic confirmation of chemical abundances lower than 
        what is found in the Sun \citep{chamberlain1951,sandage2000}. Its brightness and proximity to the Sun 
        means it has a well-determined parallax and also a well-determined age. Furthermore, it is one of the oldest 
        stars with very reliable age estimation \citep{bond2013}. 
        
        The best LTE abundance we found for HD 140283 was $A\rm{(K)}=2.95$ and the final non-LTE abundance is 
        a very low abundance of $A\rm{(K)}=2.77$ (or [K/Fe]$=-0.06$ dex), which is lower than the mean non-LTE 
        abundance we found for stars in the metallicity regime $-2.2\le$[Fe/H]$\le-2.5$ of [K/Fe]$=+0.21$ 
        dex; see Fig. \ref{gce_fig}. This lower-than-usual abundance was not expected, but the LTE abundance 
        we found matches the LTE abundance given by the radiative-transfer code MOOG \citep{sneden1973} using 
        the same stellar information.
        
        Although the best abundance is not as high as we expected for the metallicity of the star, the 
        abundance we found fits very well to the observed spectra and the resulting profile appears to be very reliable. 
        Like in HD 84937, if the best LTE abundance is plotted on top of the best non-LTE abundance the two lines 
        are almost indistinguishable, with the non-LTE line being very slightly deeper. 
        In Figure \ref{hd14} we show the non-LTE and LTE line of K with an abundance 
        of $A\rm{(K)}=2.77$.    

\subsection{HD 192263}
\label{hd19sec}

\begin{figure}[!htp]
                \centering
                \includegraphics[width=9.09cm]{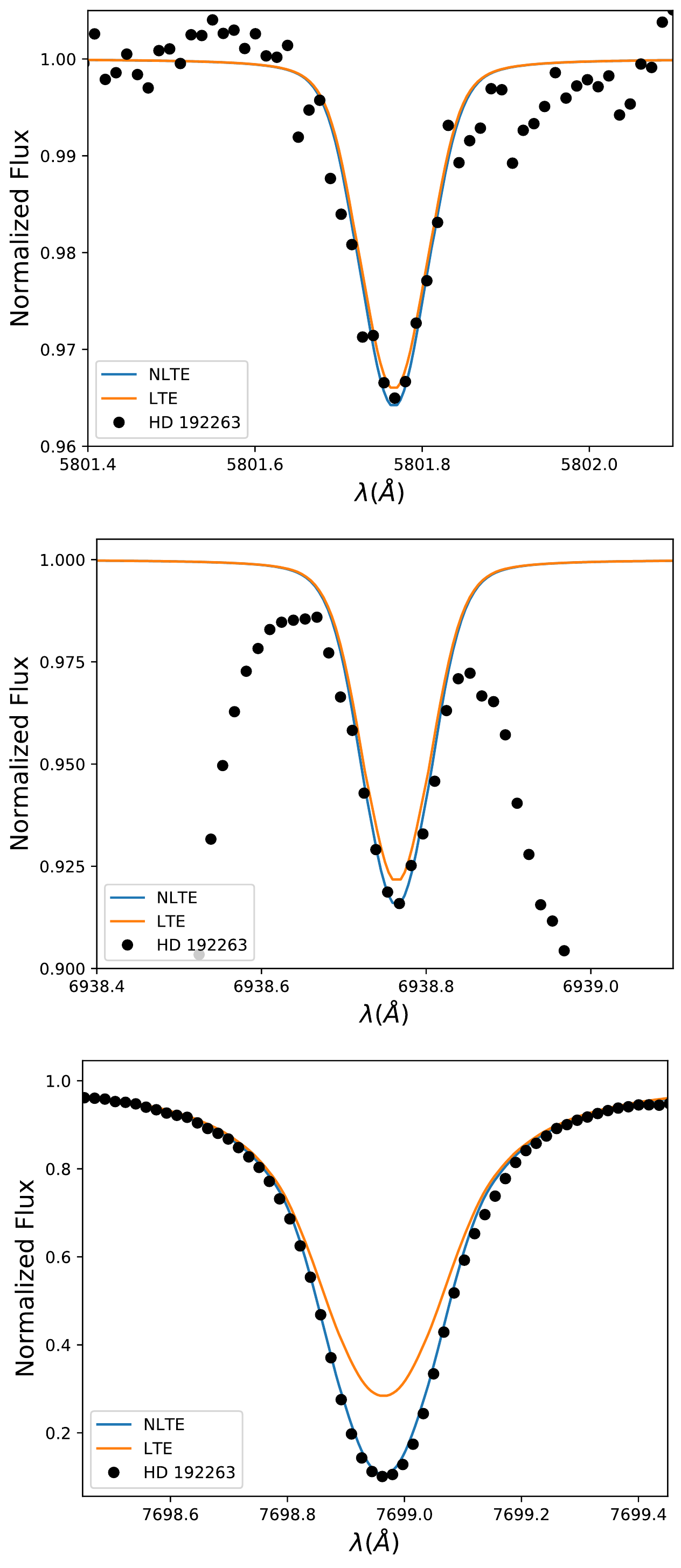}
            \caption{Non-LTE abundance fit of the $5801$, $6939,$ and $7698 \ \AA$ lines 
            (top, middle, and lower panels, respectively) 
            for HD 192263 with an abundance of $A\rm{(K)}=5.03$.  The LTE lines of the same abundance are also 
            shown in the plots.}
            \label{hd19}
\end{figure}

        HD 192263 is a cool dwarf star of nearly solar metallicity, and is also a Gaia-ESO benchmark star. 
        Under its atmospheric conditions, the $7698 \ \AA$ line develops strong damping wings and it is 
        possible to visually perceive the asymmetries at the wings of the line, as shown in Figure \ref{hd19}. 
        It is difficult to reproduce the potassium line even under non-LTE.
        
        In this star, it is possible to detect not only the resonance line, but 
        also the other two clean potassium spectral lines. We therefore determined the abundance of this star 
        by measuring the EW of the three lines ($5801 \ \AA$, $6939 \ \AA$, and $7698 \ \AA$) and 
        found the non-LTE abundances of $A\rm{(K)}=5.07$, $5.02,$ and $5.01,$ respectively. 
        The adopted value is the averaged value of $A\rm{(K)}=5.03$.
        
        The LTE abundances of the $5801 \ \AA$ and $6939 \ \AA$ lines were found to be $A\rm{(K)}=5.04$, which is 
        consistent with the non-LTE abundances. The $7698 \ \AA$ LTE abundance is $5.26$, a non-LTE 
        correction of $-0.23$ dex.

        We show the spectra and the synthetic lines of the $5801$, $6939,$ and $7698 \ \AA$ lines in Figure 
        \ref{hd19}. The observed spectra can be reproduced by the non-LTE synthetic spectra but in HD 192263 
        the synthetic line of the best LTE abundance found could not reproduce either the core or the wings 
        of the observed line. 
        Both in LTE and non-LTE there is an asymmetry at the red wing that has the same form as the 
        asymmetry observed in Procyon (Sect. \ref{procyon_sec} and
        further discussed in Sect. \ref{3dsec}).

\subsection{Procyon}
\label{procyon_sec}

\begin{figure}[!htp]
                \centering
                \includegraphics[width=9.09cm]{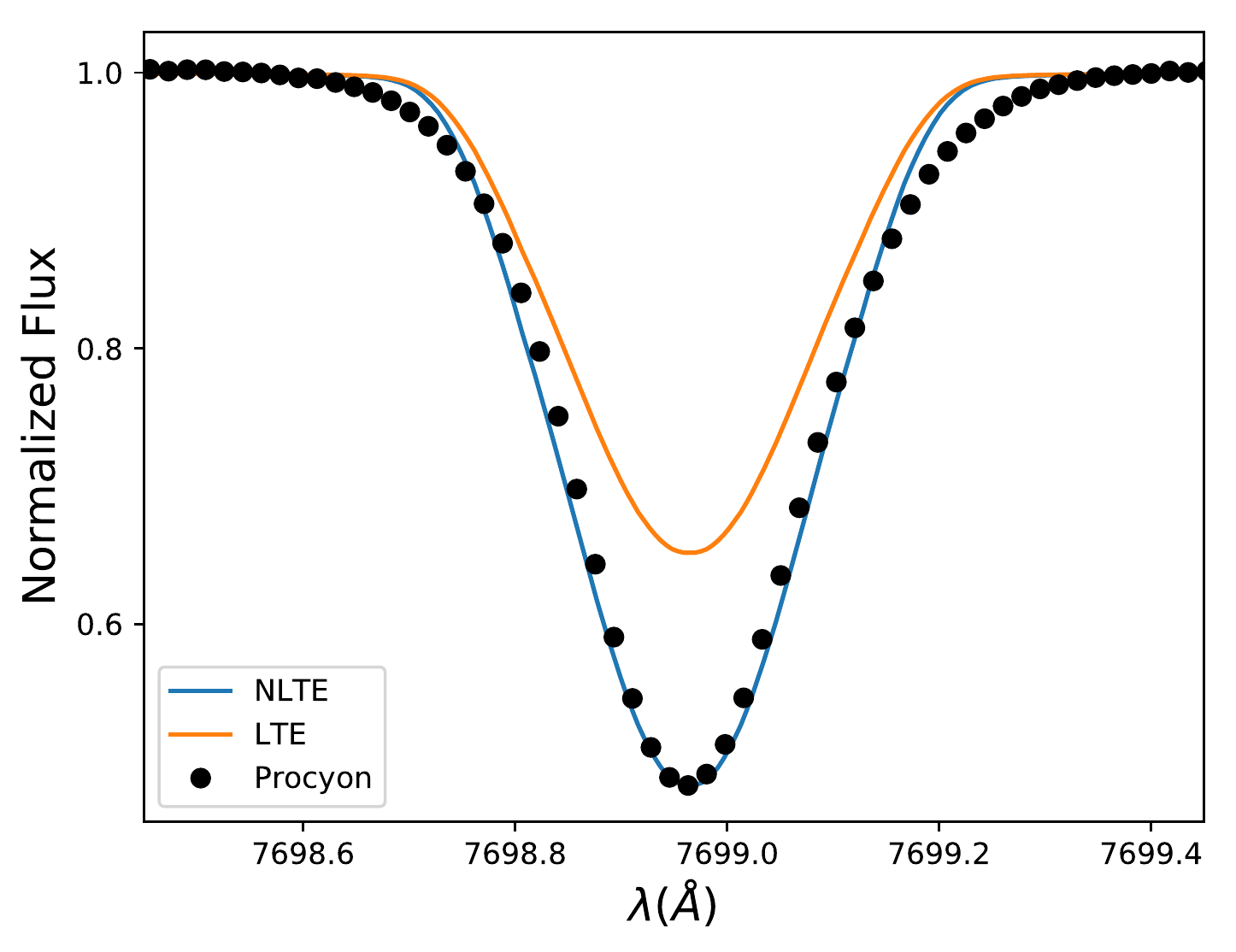}
            \caption{Non-LTE abundance fit of the $7698 \ \AA$ line for Procyon with an abundance of 
            $A\rm{(K)}=4.86$. The LTE line of the same abundance is also shown in the plot.}
            \label{proc}
\end{figure}

        Procyon is a solar metallicity F-type star very close to the Sun, 
        in which the only observable K line is the resonance line $7698 \ \AA$.
        The wings of the line are very broad and difficult to 
        correctly model. The non-LTE abundance, based on the EW of the line, fails to reproduce 
        the wings; the red wing is more broadened than the blue wing, as also observed in HD 192263, an 
        effect that is known to be a result of granulation in the stellar atmosphere \citep[e.g.,][]{dravins1981}. 
        \cite{takeda1996} 
        were able to model the wings of Procyon, but they applied corrections by hand to the red wing of the 
        line and to the line core in order to correct what they believed to be unfavorable instrumental effects in their 
        measurements.
        
        As can be seen in Figure \ref{proc} the line core is well represented by the non-LTE synthetic line, while the 
        broadened wings of the observational spectra diverge near the continuum level. We found a non-LTE abundance 
        of $A\rm{(K)}=4.86$ and a LTE abundance of $A\rm{(K)}=5.54$, a correction of $-0.68$ dex. 
        \cite{takeda1996} analyzed the potassium abundance in Procyon and also found a non-LTE correction of 
        $\approx -0.7$ dex, although their LTE and non-LTE abundances are higher than what we found.
        We note again that even in non-LTE a 1D model could not simultaneously reproduce the core and the wings of 
        the potassium $7698 \ \AA$ line in Procyon.  As before, we also show the LTE line of the same abundance 
        ($A\rm{(K)}=4.86$) in Fig. \ref{proc}.
        
\section{Three-dimensional non-LTE}
\label{3dsec}

\begin{figure}[!htp]
                \centering
                \includegraphics[width=9.09cm]{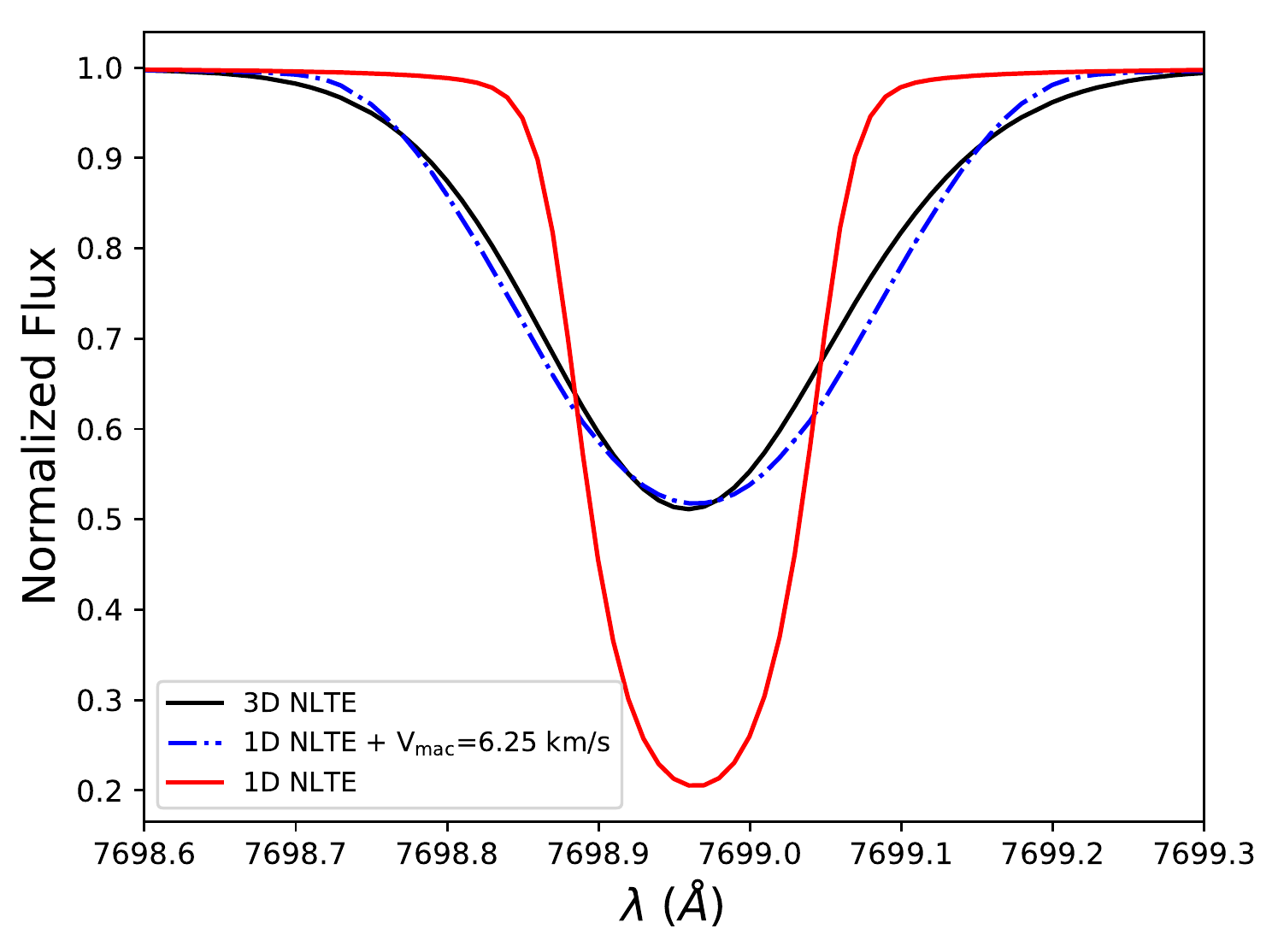}
            \caption{Synthetic non-LTE  $7698 \ \AA$ potassium line under different atmospheric model assumptions (1D and 3D).}
            \label{3dplot}
\end{figure}

In Sect. \ref{analysis} we demonstrated that there are severe (1D) non-LTE effects on the potassium 
resonance line, and that non-LTE methods are therefore needed to obtain reliable estimates of 
potassium abundances. This was particularly observed through the 
abundances of the Sun and HD 192263, in which one can measure all three optical lines, including those not 
heavily affected by non-LTE effects. In those cases only a non-LTE analysis can simultaneously give us 
consistent abundances in all observable lines. 
        
However, although we can correctly reproduce the core of the $7698 \ \AA$ line, the wings are not as 
well reproduced, an effect that is observable in the spectra of 
HD 192263 (Sect. \ref{hd10sec}) and Procyon (Sect. \ref{procyon_sec}). In Procyon one can clearly see 
the existence of an asymmetry  between the blue and red wings of the line. 
        
These types of asymmetries are associated with convection effects and can only be correctly modeled by 
using a 3D radiation-hydrodynamical simulation of the stellar atmosphere 
\citep{dravins1981,     asplund2000, asplund2004}. A great amount of effort has been invested in the modeling of 
the convection-induced asymmetry of the lithium resonance line, which is similar to potassium in many respects. 
In particular, many have attempted to disentangle the imprint of convection with possible absorption in the 
red wing due to $^6$Li \citep[e.g.,][]{smith2001, asplund2006}. The importance of accounting for 
non-LTE effects combined with the 3D line formation of lithium has also been emphasized 
\citep{cayrel2004, lind2013}.

The asymmetric shape of the potassium resonance line has been studied in the context of solar granulation 
for decades \citep[e.g.,][]{marmolino1987}. However, although there have been studies modeling the 
resonance line formation using 3D LTE models \citep[e.g.,][]{nissen2000,scott2015} there is, to the best of 
our knowledge, no published 3D non-LTE study of potassium. Although a full quantitative analysis is outside 
the scope of this project, we explored the effects with the radiative-transfer code BALDER \citep{amarsi2018}. 
We ran one snapshot of a full 3D non-LTE calculation based on a STAGGER \citep{magic2013} 
model atmosphere with $\teff=6437$ K, log $g=4.0$, [Fe/H]$=0.0,$ and [K/Fe]$=0.0$, which correspond 
to similar      parameters to those of Procyon.
        
Our results showed that 1D non-LTE can only partially reproduce the wings seen in 3D 
after we added strong macroturbulent broadening effect of V$_{\rm{MAC}}$ $\approx 6.25$ km/s (dashed line), 
and even so the 1D non-LTE does not fully reproduce the wings seen in 3D, 
particularly the asymmetry of the line. 
                
The 3D non-LTE feature that we synthesized indicates that the misrepresentation of the wings 
in our 1D non-LTE  analysis is due to unaccounted-for 3D effects. Although the asymmetries of the 
potassium resonance wings are much better represented in full 3D non-LTE we argue that the EW analysis of the 1D non-LTE yields comparable results. 

\section{Non-LTE corrections grid}
\label{grid}

\begin{figure}[!htp]
                \centering
                \includegraphics[width=9.09cm]{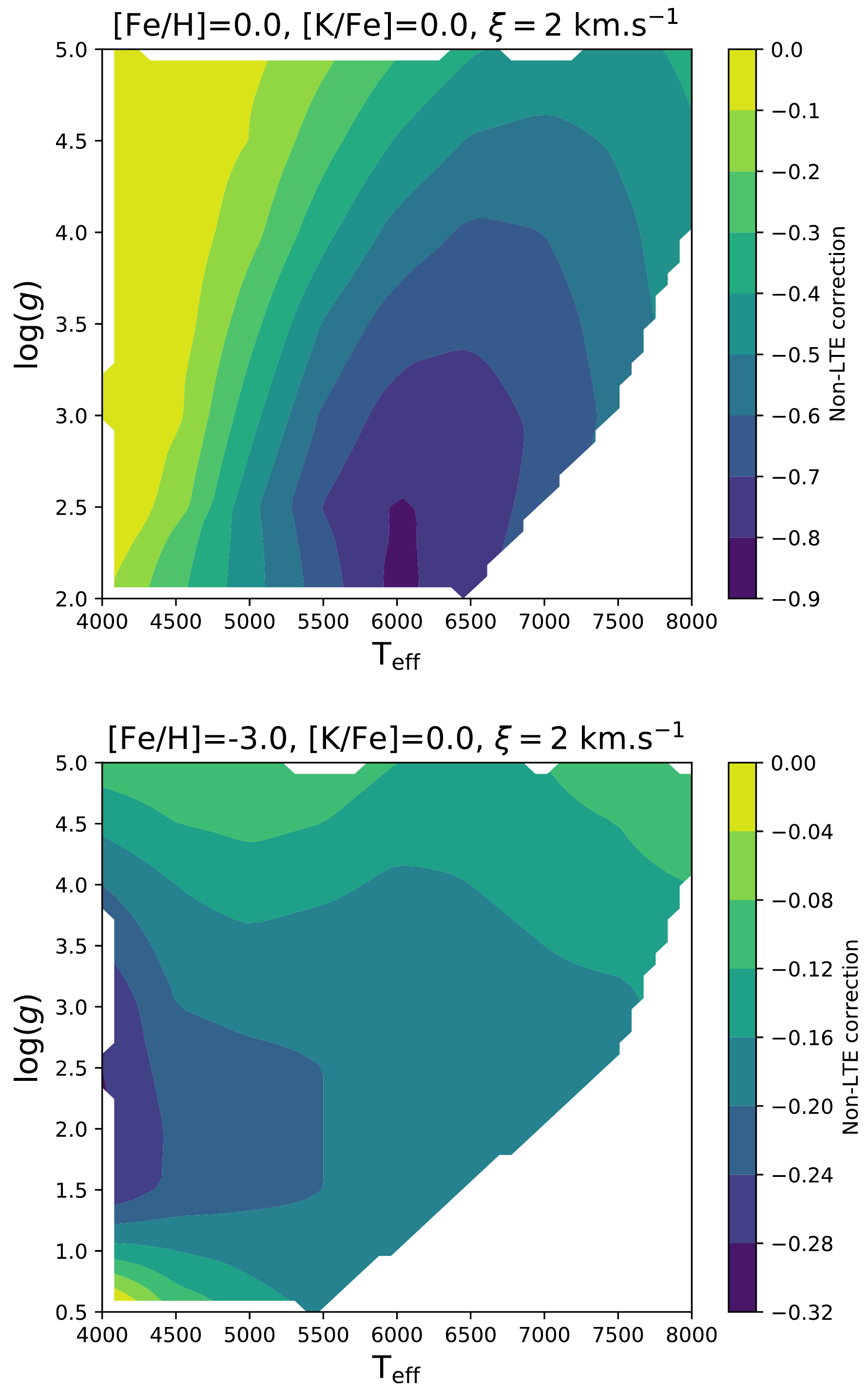}
            \caption{Contour diagram illustrating the abundance corrections in [Fe/H]$=0.0$ (upper panel) and 
            [Fe/H]$=-3.0$ (lower panel) for the $7698 \ \AA$ line.}
            \label{contour}
\end{figure}

\begin{figure}[!htp]
                \centering
                \includegraphics[width=9.09cm]{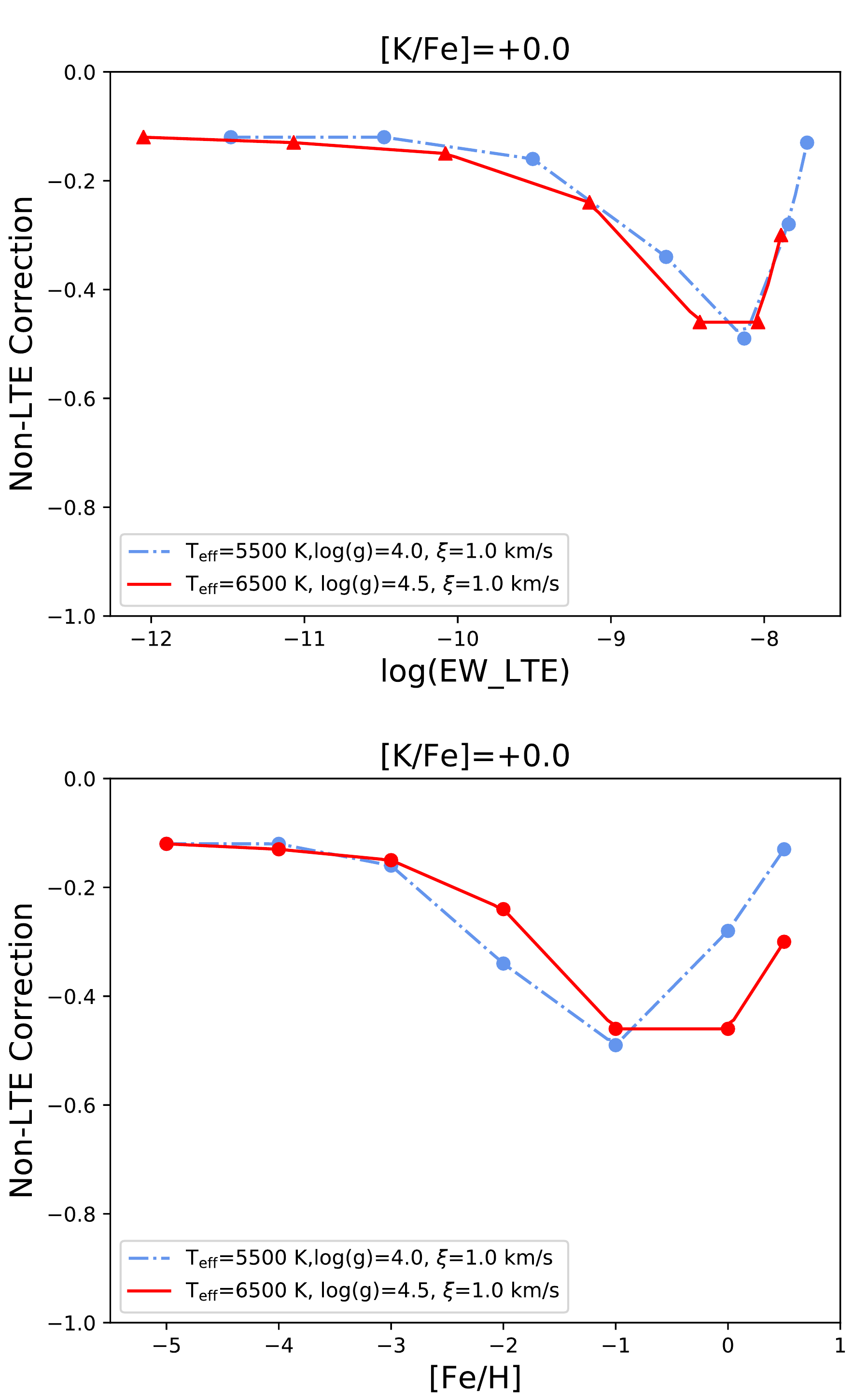}
            \caption{The top panel shows how the non-LTE corrections vary with EW. The lower 
            panel shows the non-LTE correction variation with [Fe/H]. In both panels we have two different 
            model atmospheres and both show the corrections for the $7698 \ \AA$ line.}
            \label{corrxfeh}
\end{figure}

Using our standard atom we produced a grid of non-LTE  corrections, for model atmospheres of 
different stellar parameters. Our grid was computed for models with 
effective temperatures in the range $4000\le  \rm{T}_{\rm{eff}} \   / \rm{K}  \le 8000$  
with steps of $500$ K; for each T$_{\rm{eff}}$ the surface gravity range is  $0.5 \le \rm{log}(g) \le5.0$ dex in 
steps of $0.5$, and we also vary the metallicities in the range $-5.00 \le$ [Fe/H] $+0.50$ in steps of $0.25$ and 
use microturbulence velocities of $\xi=1.0$, $2.0,$ and $5.0$ $\rm{km/s}$. 
We determined synthetic spectral lines of potassium abundances varying in the range $-1.25\le$[K/Fe] $\le+1.25$, 
and estimated the LTE and non-LTE  EWs in each case. Thus, we obtained a final grid 
with synthetic LTE and non-LTE  EWs for each calculated abundance in each model atmosphere.

In Figure \ref{contour} we show an example of the non-LTE corrections for solar metallicity varying the stellar 
parameters. The abundance correction for a Sun-like star is approximately $-0.3$ dex and for stars like Procyon the 
abundance correction is as high as $-0.7$ dex. 

From the top panel of Figure \ref{corrxfeh} we can see that the correction is very dependent on line 
strength, and the apparent dependence with temperature seen in Fig. \ref{contour} is an indirect 
effect. 
The extremely high non-LTE corrections for certain stellar atmospheres shows us that it is imperative to apply 
non-LTE corrections when analyzing the abundances of potassium from the resonance lines.

In the lower panel of Figure \ref{corrxfeh} we show the dependence with metallicity of two model atmospheres 
for a K abundance of [K/Fe]=0.0. An important dependence with metallicity starts to appear at [Fe/H] greater than 
$-2.0$; this can be directly related to the line becoming stronger for higher metallicities and saturating.
As the metallicity increases and the line gets stronger, the corrections are larger and peak at approximately solar 
metallicity before decreasing again. However, we have not found any measurements of 
potassium in stars with metallicities higher than solar, and therefore this decrease in the corrections after [Fe/H]=0.0 
has not been applied in any observational data to test its compatibility with the GCE models. 

We also computed the corrections for the resonance $7664 \ \AA$ line and the infrared lines $15163 \ \AA$ 
and $15168 \ \AA$ (which are in the APOGEE range). 
Abundances from the resonance $7664 \ \AA$ line must also be non-LTE corrected as the corrections are on the 
same order as the corrections computed for the $7698 \ \AA$ line, but the infrared lines do not have such 
strong non-LTE dependence (the mean correction of the $15168 \ \AA$ for [K/Fe]$=0.0$ at the same models 
presented in Figure \ref{corrxfeh} are $-0.03$ and $-0.06$ dex).

Our final grid is publicly available as an electronic table.
\section{Chemical evolution of Potassium}
\label{gce}

\subsection{Metal-poor stars}

        We applied our grid of corrections 
        to a series of abundances from several publications in order to study the evolution of potassium in our 
        galaxy. We show these results and compare them to the K11
        \citep{kobayashi2011}, K15 \citep{zhao2016,sneden2016} GCE models, and 
        the models from \cite{prantzos2018} with and without yields from massive rotating stars.

        We determined abundances under LTE using the 1D LTE code MOOG \citep{sneden1973} based on  
        three sets of published EWs: \cite{cayrel2004}, \cite{roederer2014}, and \cite{spina2016}. 
        We then determined the non-LTE abundances by applying non-LTE abundance corrections  
        from our grid. \cite{andrievsky2010} recently reanalyzed the \cite{cayrel2004} sample with their 
        non-LTE corrections and we compared our results to theirs, finding a mean difference of $\sim -0.1$ dex 
        in the final non-LTE abundances (with their abundances being higher).
        
        Figure \ref{gce_fig} and Table \ref{nlte_cor_abnd_table} show both the [K/Fe]$_{\rm{LTE}}$ and the [K/Fe]$_{\rm{NLTE}}$ results for the above-mentioned 
        samples. 
        In the upper panel of Fig. 14 we see the behavior of the LTE abundances. 
        For all data there is a large discrepancy between the abundances of models and observations, 
        reaching up to $1$ dex. For [Fe/H]$\lesssim-1.0$, the LTE potassium abundances show an 
        increasing trend with [Fe/H], which is  similar to what is predicted by the 
        \cite{kobayashi2011} model, but differs from the behavior of the \cite{prantzos2018} models. 
                
        The non-LTE  abundances however show a different result. In the region [Fe/H]$\lesssim-1.0$, there is no 
        longer a trend of increasing potassium abundances with increasing [Fe/H]; rather, the potassium abundances 
        gradually decrease. The model of  rotating massive stars by 
        \cite{prantzos2018} appears to correctly reproduce the observations, although the mean observed abundance 
        is slightly higher than the model. Non-LTE corrected abundances and the  model of massive rotating stars show  
        the same behavior: a small increase in abundances between $-3\le$ [Fe/H] $\le -2$ followed by a 
        decrease in abundances with increasing metallicity. On the other hand, the models from \cite{kobayashi2011} 
        and the \cite{prantzos2018} model without rotation underestimate the production 
        of potassium and their yields clearly fail to reproduce the observations.

\subsection{Solar twins}

\begin{figure}[!htp]
                \centering
                \includegraphics[width=9.09cm]{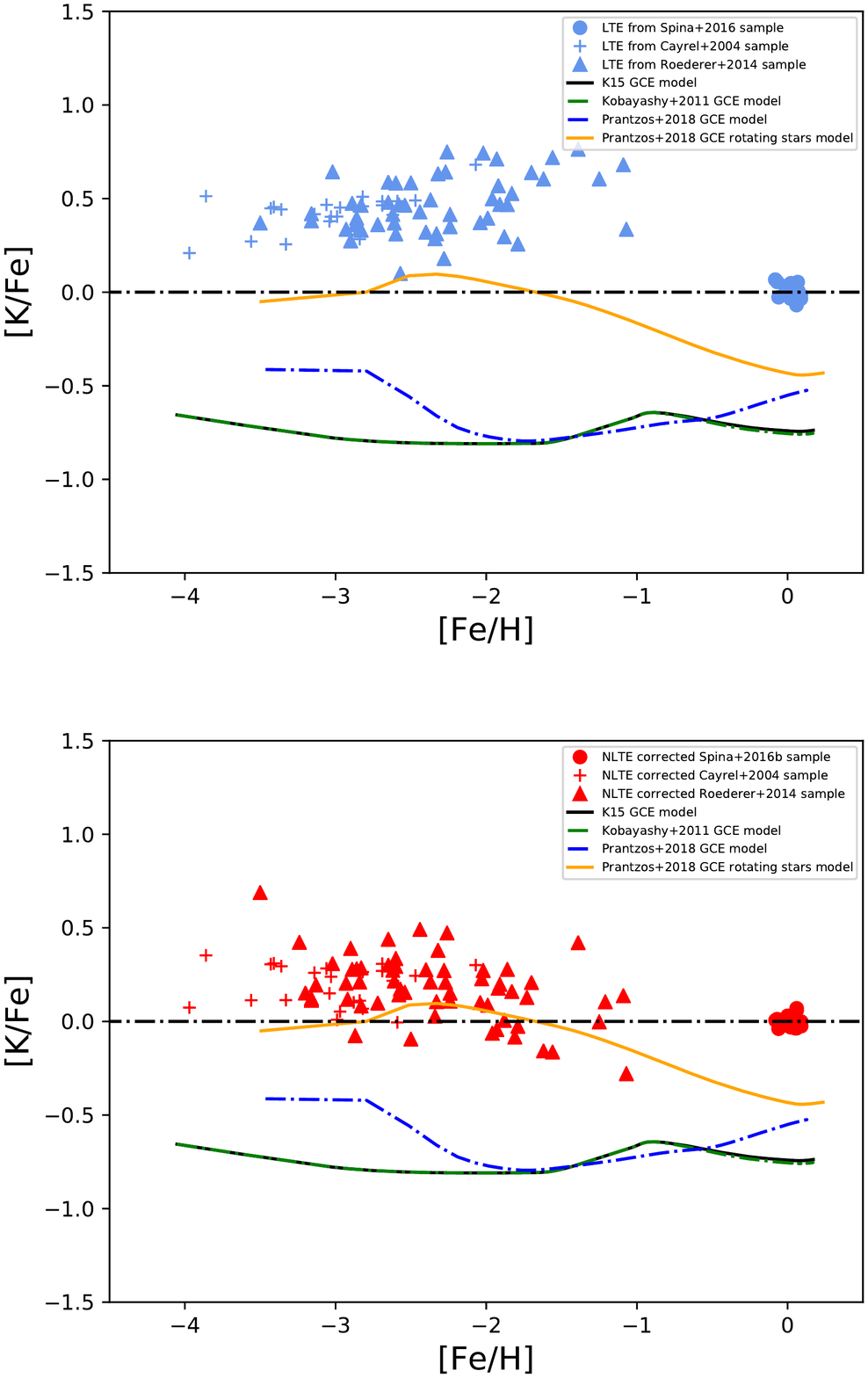}
            \caption{(Top panel) LTE abundances calculated using MOOG and the EWs from 
            \cite{cayrel2004}, \cite{roederer2014}, and \cite{spina2016}. (Lower panel) Non-LTE corrected abundances 
            (from the LTE abundances in the upper panel). The LTE and non-LTE corrected abundances (in A(K)), are shown 
            in Table \ref{nlte_cor_abnd_table}. In both panels we also show three different GCE models: 
            in black the model from \cite{kobayashi2011}; in dashed blue and orange the models from 
            \cite{prantzos2018}  without and with yields from massive rotating stars, respectively. }
            \label{gce_fig}
\end{figure}

        The solar twin sample of Spina et al. (2016) sits at [Fe/H]$\approx0.0$ in Fig. \ref{gce_fig}. Correcting the 
        abundances for non-LTE effects does not have a significant effect here. This is because these abundances 
        were measured in a line-by-line differential analysis with respect to the Sun. To first order, the non-LTE errors 
        in the analysis of the solar twin spectra cancel with the non-LTE errors in the analysis of the solar spectrum. 
        However,  the non-LTE corrections lead to a reduction in the scatter of the potassium abundances, as we 
        discuss below.  

\begin{figure}[!htp]
                \centering
                \includegraphics[width=9.09cm]{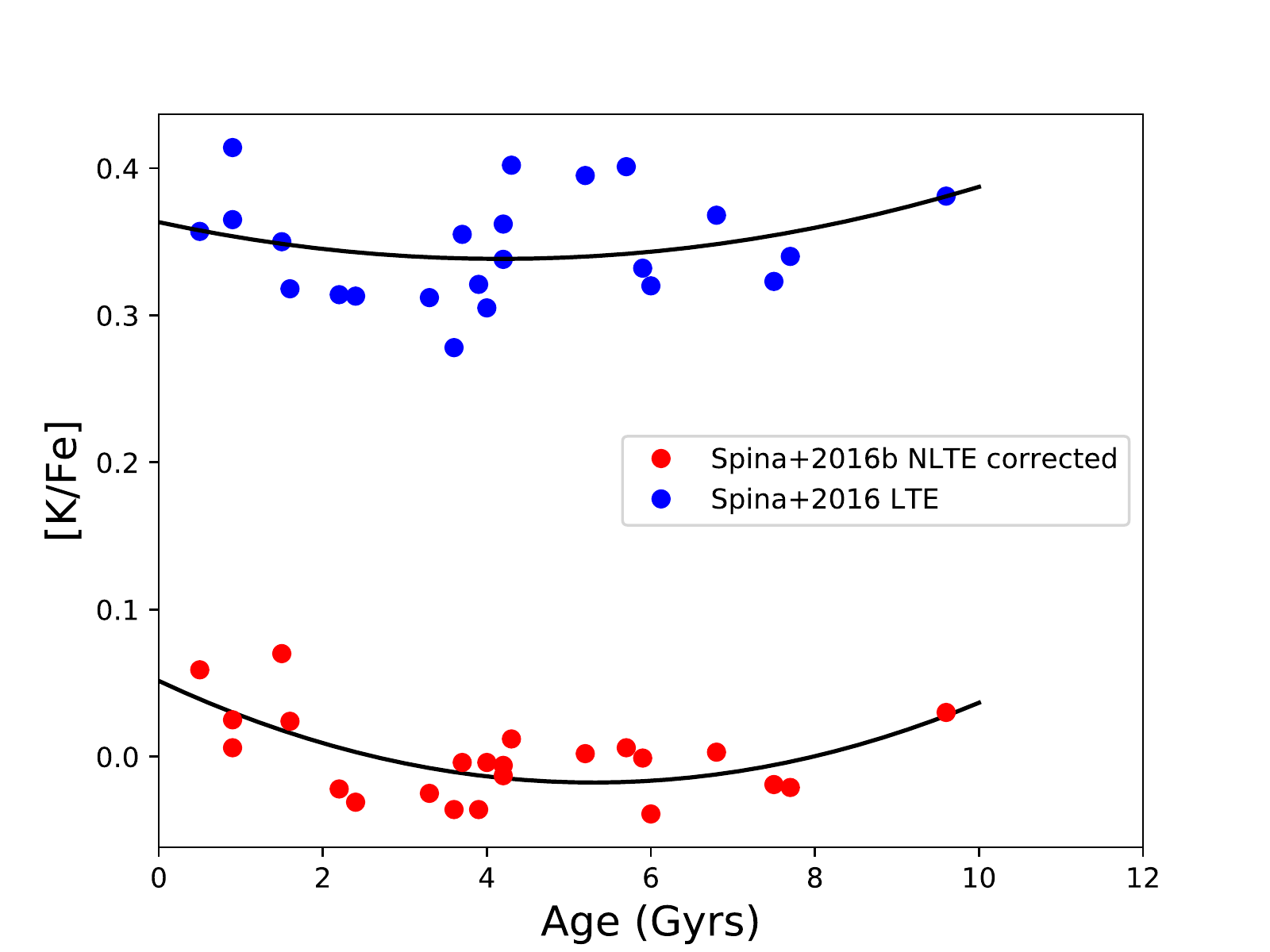}
            \caption{In red we show the non-LTE  abundances of the \cite{spina2016} sample and in light blue the 
            scaled LTE abundances. The offset is due to the use of the same solar potassium abundance 
            LTE and non-LTE.}
            \label{solar_twins}
\end{figure}

        In their high-precision differential analysis of solar twins (stars with atmospheric parameters similar to the 
        solar parameters  - $\Delta \rm{T}_{\rm{eff}} \pm 100 $ K, $\Delta \rm{log} (g)\pm 0.1$ dex, 
        $\Delta \rm{[Fe/H]} \pm 0.1$ dex and mass within $\approx 5 \%$ of the solar mass) \cite{spina2016} 
        analyzed 20 elements and for all of them, except potassium, they found a tight relationship between stellar age 
        and [X/Fe], with strongly varying slope depending on the element. With this in mind 
        we reanalyzed their sample of solar twins and the same plot of [K/Fe] versus age can be seen in Fig. 
        \ref{solar_twins}. The trends with stellar age of the LTE and the non-LTE  abundances are the same. Being 
        solar twins, not only are the abundances very similar to one another, but also the non-LTE corrections. 
        Although this result was expected, we call attention to the fact that the non-LTE abundances have a smaller 
        scatter than the LTE abundances. This means that the non-LTE abundances might be more reliable, as the 
        abundances of solar twins are very homogeneous and similar to the solar abundance 
        \citep{bedell2018}. 
        This is also important because it shows that even in a differential abundance analysis of solar twins,
        non-LTE corrections 
        play an important role in explaining the evolution of the elements by decreasing the scatter of 
        abundance measurements, which might further improve fits of their behavior. The reduced 
        scatter is important to correctly assess the homogeneity of chemical evolution and therefore to draw more precise 
        conclusions on the nucleosynthetic mechanisms that drive potassium evolution throughout cosmic history. 
        The reduced scatter is also important to assess the homogeneity of different stellar populations 
        (e.g., thin/thick disk) and 
        stellar associations (clusters or binaries), which is also relevant for chemical tagging, a popular  
        application of stellar spectroscopy.
        
        As already mentioned, \cite{spina2016} found a function able to describe [X/Fe] abundances over 
        cosmic time for all elements analyzed, apart from potassium. 
        We also tried to find a function to describe [K/Fe] over time 
        with our lower scatter non-LTE abundances, but as in \cite{spina2016} we did not find a function that 
        can describe its behavior. Figure \ref{solar_twins} shows a fourth-degree 
        polynomial function that was used only to demonstrate the smaller scatter of the non-LTE 
        data. The standard deviations of the non-LTE-corrected and the LTE abundances are $0.027$ dex and 
        $0.035$ dex, respectively, an improvement of $\sim 23\%$.
        Within $1\sigma$ the LTE and non-LTE fits are the same.
        
        It is interesting how there is no possible fit to the potassium abundances over time, as found for the other 
        elements. This might indicate an inhomogeneous enrichment of potassium, although the small scatter 
        we observe might be caused by errors in the analysis. If indeed the ISM was not fully homogenized with K, the
        small abundance differences among stars born in different sites in the Galaxy could explain this uncanny inability 
        to fit the abundances. This behavior could be explained by a GCE where rotating massive stars play a role. 
        As the occurrence of such objects is not as common as, for example, type II supernovae, 
        rotating massive stars might be able to 
        produce these localized small inhomogeneities, which would depend on a localized mass function and the 
        production of the rotating massive stars. This localized effect would not be seen in most 
        of the other commonly measured elements (such as C, O, Na, Mg, Al, Si, etc) because, 
        as one can see in Fig. 13 of \cite{prantzos2018}, most of them have an evolution that is not significantly  
        influenced by massive rotating stars, as both models (with and without the massive rotating stars) have 
        virtually the same results. However, we caution that such small scatter might also be caused by errors in our analysis, and that the inability to fit the solar-twin data might  simply be because the empirical fits suggested are not 
        adequate.

\section{Conclusions}
\label{conclusao}

We built a new model atom of potassium in order to test the non-LTE effects in stellar abundances estimations. 
We used state-of-the-art calculations of neutral hydrogen and electronic collisional 
cross-sections and photoionization cross-sections. 

The use of improved calculations of electron collisions were the most influential factor 
in the non-LTE line profile of the late-type stars and both CCC and BSR calculations give comparable results. 
The choice between the different H+K collisions and the DBSR photoionization cross-sections also had 
non-negligible and similar effects on the final line 
depth of the synthetic lines, and their contribution is more relevant to the analysis of giant stars. 
The use of different collisional recipes can lead to differences 
of up to $0.1$ dex in the abundance analysis. With the aid of an atom with only the resonance transitions 
we concluded that the major driver of the non-LTE effects is the photon losses in the $7664$ and $7698 \ \AA$ 
lines.

 As
test cases for the quality of 
our synthetic spectra, we derived the potassium abundance of the Sun and other benchmark stars. We found that our non-LTE  synthetic spectra are a good fit to the stellar spectra, as 
opposed to what is seen in LTE. When available the abundances were derived from the three optical 
spectral lines and the results are very homogeneous with the non-LTE  approach, and vary considerably if
every line is analyzed in LTE. 
Through this analysis we also showed that there are important 3D effects that can be observed 
via the asymmetries in the wings of the potassium line. The differences of the 1D and 3D line profiles were 
shown for a test case and we see that EW analyses of 1D non-LTE  and 3D non-LTE  lines 
give similar results, which is valuable to validate the grid of corrections that was developed for this work.

Finally, we used our grid of corrections on three sets of potassium data to analyze the chemical evolution 
of the Galaxy. We see that the yields of the common models (Figure \ref{gce_fig}) 
are not enough to explain the observational data. However, the model with yields from 
rotating massive stars by \cite{prantzos2018} is able to partially reproduce the core of the 
observed abundances, especially the decrease in abundance for metallicities higher 
than [Fe/H] of around $-2.5$ dex, which we also observe in our non-LTE  results from the 
\cite{roederer2014} sample. All the GCE models fail to reproduce the observed abundances at solar metallicity.
However, the decreased scatter in the solar twins is an important result as it shows that the LTE 
differential abundance is not fully free of systematic uncertainties due to line formation responsible 
for part of the observed inhomogeneities, and will help constrain the nucleosynthetic mechanism 
mainly responsible for potassium production in this narrow metallicity range. 

\begin{acknowledgements}
        We thank the referee for their careful reading of the manuscript, which helped us to improve its  quality.
        HR thanks his PDSE program (88881.132145/2016-01). This study was financed in part by the 
        Coordenação de Aperfeiçoamento de Pessoal de Nível Superior - Brasil (CAPES) - Finance Code 001. AMA and KL 
        acknowledge funds from the Alexander von Humboldt Foundation in the framework of the Sofja Kovalevskaja 
        Award endowed by the Federal Ministry of Education and Research, and KL also acknowledges funds from 
        the Swedish Research Council (grant 2015-004153) and Marie Sk{\l}odowska Curie Actions 
        (cofund project INCA 600398). PSB acknowledges financial support from the Swedish Research Council and 
        the project grant ``The New Milky Way'' from the Knut and Alice Wallenberg Foundation. The work of OZ and KB 
        was supported by the United States National Science Foundation under grant 
        Nos.\ PHY-1403245, PHY-1520970, and PHY-1803844. The numerical BSR calculations were performed 
        on STAMPEDE at the Texas Advanced Computing Center. They were made possible by the XSEDE 
        allocation No.\ PHY-090031.  L.S. is supported from the Australian Research Council (Discovery Project 
        170100521). JM thanks support of FAPESP (2014/18100-4, 2018/04055-8) and CNPq (Bolsa de 
         Produtividade). This work has made use of the VALD database, operated at Uppsala University, the Institute 
         of Astronomy RAS in Moscow, and the University of Vienna.
\end{acknowledgements}

%
%

\begin{appendix}
\section{Non-LTE corrected abundances}

\longtab[1]{
\begin{longtable}{lrrrrrr}
  \caption{\label{nlte_cor_abnd_table}Non-LTE corrected abundances. Stellar parameters and LTE 
  abundances are from   \cite{spina2016,cayrel2004} and \cite{roederer2014}}\\
        \hline \hline
        Star & T$_{\rm{eff}}$ (K) & $\log \ g$ ($g$ in $\rm{cm.s^{-2}}$) & [Fe/H] & $\xi$ ($\rm{km.s^{-1}}$) & K$_{\rm{NLTE}}$ & K$_{\rm{LTE}}$ \\ 
      \hline    
      \endfirsthead
      \caption{Continued}\\
      \hline\hline
        Star & T$_{\rm{eff}}$ (K) & $\log \ g$ ($g$ in $\rm{cm.s^{-2}}$) & [Fe/H] & $\xi$ ($\rm{km.s^{-1}}$) & K$_{\rm{NLTE}}$ & K$_{\rm{LTE}}$ \\ 
\hline
\endhead
\hline
\endfoot
\hline
\endlastfoot
18Sco & $5818$ & $4.46$ & $0.05$ & $1.00$ & $5.14$ & $5.48$ \\ 
HIP 56948 & $5793$ & $4.47$ & $0.02$ & $1.00$ & $5.10$ & $5.44$ \\ 
HIP 102152 & $5724$ & $4.36$ & $-0.01$ & $1.00$ & $5.08$ & $5.44$ \\ 
HD 20630 & $5744$ & $4.49$ & $0.06$ & $1.10$ & $5.23$ & $5.53$ \\ 
HD 202628 & $5831$ & $4.51$ & $0.00$ & $1.00$ & $5.13$ & $5.47$ \\ 
HIP 30502 & $5736$ & $4.42$ & $-0.06$ & $1.00$ & $5.01$ & $5.37$ \\ 
HIP 73815 & $5800$ & $4.33$ & $0.03$ & $1.10$ & $5.14$ & $5.50$ \\ 
HIP 77883 & $5701$ & $4.37$ & $0.02$ & $1.00$ & $5.11$ & $5.45$ \\ 
HIP 89650 & $5853$ & $4.42$ & $-0.01$ & $1.10$ & $5.09$ & $5.46$ \\ 
HD 9986 & $5827$ & $4.44$ & $0.09$ & $1.00$ & $5.17$ & $5.51$ \\ 
HD 13531 & $5653$ & $4.53$ & $0.02$ & $1.20$ & $5.15$ & $5.45$ \\ 
HD 13931 & $5895$ & $4.29$ & $0.07$ & $1.10$ & $5.19$ & $5.58$ \\ 
HD 32963 & $5768$ & $4.37$ & $0.09$ & $1.00$ & $5.20$ & $5.53$ \\ 
HD 33636 & $5963$ & $4.47$ & $-0.08$ & $1.10$ & $5.04$ & $5.44$ \\ 
HD 43162 & $5661$ & $4.53$ & $0.06$ & $1.20$ & $5.24$ & $5.52$ \\ 
HD 45184 & $5873$ & $4.41$ & $0.07$ & $1.00$ & $5.18$ & $5.54$ \\ 
HD 87359 & $5700$ & $4.47$ & $0.07$ & $0.90$ & $5.18$ & $5.48$ \\ 
HD 95128 & $5904$ & $4.35$ & $0.02$ & $1.10$ & $5.13$ & $5.53$ \\ 
HD 98618 & $5845$ & $4.42$ & $0.05$ & $1.10$ & $5.12$ & $5.48$ \\ 
HD 106252 & $5885$ & $4.42$ & $-0.07$ & $1.10$ & $5.05$ & $5.44$ \\ 
HD 140538 & $5704$ & $4.48$ & $0.06$ & $0.90$ & $5.13$ & $5.45$ \\ 
HD 143436 & $5825$ & $4.43$ & $0.04$ & $1.00$ & $5.14$ & $5.49$ \\ 
HD 112257 & $5686$ & $4.30$ & $-0.00$ & $0.90$ & $5.14$ & $5.49$ \\ 
HD 2796 & $4950$ & $1.50$ & $-2.47$ & $2.10$ & $2.88$ & $3.13$ \\ 
HD 122563 & $4600$ & $1.10$ & $-2.82$ & $2.00$ & $2.36$ & $2.75$ \\ 
HD 186478 & $4700$ & $1.30$ & $-2.59$ & $2.00$ & $2.52$ & $3.01$ \\ 
BD +17 3248 & $5250$ & $1.40$ & $-2.07$ & $1.50$ & $3.34$ & $3.72$ \\ 
BD -18 5550 & $4750$ & $1.40$ & $-3.06$ & $1.80$ & $2.33$ & $2.52$ \\ 
BS 16477-003 & $4900$ & $1.70$ & $-3.36$ & $1.80$ & $2.04$ & $2.19$ \\ 
BS 17569-049 & $4700$ & $1.20$ & $-2.88$ & $1.90$ & $2.33$ & $2.69$ \\ 
CS 22169-035 & $4700$ & $1.20$ & $-3.04$ & $2.20$ & $2.22$ & $2.45$ \\ 
CS 22172-002 & $4800$ & $1.30$ & $-3.86$ & $2.20$ & $1.60$ & $1.76$ \\ 
CS 22873-055 & $4550$ & $0.70$ & $-2.99$ & $2.20$ & $2.13$ & $2.52$ \\ 
CS 22873-166 & $4550$ & $0.90$ & $-2.97$ & $2.10$ & $2.19$ & $2.59$ \\ 
CS 22892-052 & $4850$ & $1.60$ & $-3.03$ & $1.90$ & $2.32$ & $2.48$ \\ 
CS 22896-154 & $5250$ & $2.70$ & $-2.69$ & $1.20$ & $2.69$ & $2.88$ \\ 
CS 22897-008 & $4900$ & $1.70$ & $-3.41$ & $2.00$ & $2.01$ & $2.15$ \\ 
CS 22948-066 & $5100$ & $1.80$ & $-3.14$ & $2.00$ & $2.23$ & $2.39$ \\ 
CS 22949-037 & $4900$ & $1.50$ & $-3.97$ & $1.80$ & $1.21$ & $1.35$ \\ 
CS 22952-015 & $4800$ & $1.30$ & $-3.43$ & $2.10$ & $1.99$ & $2.13$ \\ 
CS 22953-003 & $5100$ & $2.30$ & $-2.84$ & $1.70$ & $2.38$ & $2.55$ \\ 
CS 22956-050 & $4900$ & $1.70$ & $-3.33$ & $1.80$ & $1.89$ & $2.04$ \\ 
CS 22966-057 & $5300$ & $2.20$ & $-2.62$ & $1.40$ & $2.71$ & $2.90$ \\ 
CS 22968-014 & $4850$ & $1.70$ & $-3.56$ & $1.90$ & $1.66$ & $1.82$ \\ 
CS 29495-041 & $4800$ & $1.50$ & $-2.82$ & $1.80$ & $2.55$ & $2.80$ \\ 
CS 29518-051 & $5200$ & $2.60$ & $-2.69$ & $1.40$ & $2.73$ & $2.91$ \\ 
HD 94028 & $5730$ & $3.70$ & $-1.81$ & $1.00$ & $3.22$ & $4.20$ \\ 
HD 175305 & $4920$ & $2.30$ & $-1.56$ & $1.40$ & $3.39$ & $4.27$ \\ 
HD 11582 & $5020$ & $2.20$ & $-2.03$ & $1.40$ & $3.31$ & $4.08$ \\ 
BD +29 2356 & $4710$ & $1.75$ & $-1.62$ & $1.50$ & $3.33$ & $4.09$ \\ 
HD 201891 & $5840$ & $4.10$ & $-1.21$ & $1.10$ & $4.00$ & $4.72$ \\ 
HD 45282 & $5230$ & $2.90$ & $-1.73$ & $1.40$ & $3.51$ & $4.21$ \\ 
G 161-073 & $5680$ & $3.90$ & $-1.07$ & $1.10$ & $3.76$ & $4.38$ \\ 
BD +19 1185A & $5440$ & $4.30$ & $-1.25$ & $1.10$ & $3.86$ & $4.46$ \\ 
HD 126238 & $4750$ & $1.65$ & $-1.96$ & $1.60$ & $3.09$ & $3.65$ \\ 
G 153-064 & $5630$ & $4.15$ & $-1.09$ & $0.90$ & $4.16$ & $4.70$ \\ 
CS 22873-128 & $4710$ & $1.20$ & $-3.24$ & $1.60$ & $2.29$ & $2.78$ \\ 
HD 178443 & $5170$ & $1.45$ & $-2.02$ & $1.90$ & $3.36$ & $3.83$ \\ 
BD -15 5781 & $4550$ & $0.70$ & $-2.87$ & $1.70$ & $2.16$ & $2.60$ \\ 
G 126-062 & $5970$ & $3.85$ & $-1.70$ & $1.10$ & $3.62$ & $4.05$ \\ 
HD 31128 & $5630$ & $3.60$ & $-1.92$ & $1.10$ & $3.37$ & $3.76$ \\ 
HD 219617 & $5730$ & $3.70$ & $-1.83$ & $1.30$ & $3.44$ & $3.81$ \\ 
CS 22955-110 & $6710$ & $3.45$ & $-1.39$ & $1.90$ & $4.14$ & $4.48$ \\ 
CS 22185-007 & $4730$ & $1.30$ & $-3.02$ & $1.60$ & $2.40$ & $2.73$ \\ 
CS 22182-047 & $5640$ & $3.60$ & $-1.99$ & $1.10$ & $3.21$ & $3.52$ \\ 
CS 22892-052 & $4690$ & $1.15$ & $-3.16$ & $1.50$ & $2.07$ & $2.37$ \\ 
HD 188510 & $5210$ & $4.10$ & $-1.88$ & $0.80$ & $3.23$ & $3.52$ \\ 
BD -01 2582 & $4920$ & $1.80$ & $-2.65$ & $1.50$ & $2.76$ & $3.05$ \\ 
HD 122196 & $5880$ & $3.80$ & $-1.79$ & $1.20$ & $3.29$ & $3.58$ \\ 
CS 29495-005 & $5990$ & $3.75$ & $-2.26$ & $1.10$ & $3.32$ & $3.60$ \\ 
CS 29513-032 & $6080$ & $3.85$ & $-1.91$ & $1.30$ & $3.40$ & $3.67$ \\ 
G 090-025 & $5150$ & $4.05$ & $-2.04$ & $0.90$ & $3.17$ & $3.44$ \\ 
CS 30492-110 & $4660$ & $1.05$ & $-3.16$ & $1.80$ & $2.06$ & $2.33$ \\ 
CS 29514-017 & $5270$ & $2.80$ & $-2.34$ & $1.20$ & $2.80$ & $3.06$ \\ 
CS 29513-014 & $5440$ & $1.55$ & $-2.32$ & $2.00$ & $3.17$ & $3.42$ \\ 
HD 108317 & $5030$ & $2.10$ & $-2.60$ & $1.40$ & $2.85$ & $3.09$ \\ 
G 090-003 & $5680$ & $3.60$ & $-2.24$ & $1.20$ & $2.98$ & $3.22$ \\ 
G 025-024 & $5670$ & $3.55$ & $-2.28$ & $1.70$ & $3.10$ & $3.34$ \\ 
HD 119516 & $5660$ & $1.90$ & $-1.93$ & $1.90$ & $3.14$ & $3.36$ \\ 
CS 22943-095 & $6140$ & $3.80$ & $-2.44$ & $1.40$ & $3.16$ & $3.38$ \\ 
CS 22881-036 & $5940$ & $3.70$ & $-2.37$ & $1.10$ & $2.95$ & $3.17$ \\ 
HD 19445 & $5820$ & $3.65$ & $-2.40$ & $1.20$ & $2.98$ & $3.20$ \\ 
CS 22968-026 & $5850$ & $3.65$ & $-2.33$ & $1.20$ & $2.88$ & $3.10$ \\ 
HD 13979 & $4830$ & $1.60$ & $-2.72$ & $1.60$ & $2.49$ & $2.70$ \\ 
CS 22958-065 & $6020$ & $3.75$ & $-2.24$ & $1.30$ & $3.02$ & $3.23$ \\ 
CS 22884-020 & $6040$ & $3.75$ & $-2.27$ & $1.40$ & $3.05$ & $3.25$ \\ 
CS 22894-004 & $5920$ & $3.65$ & $-2.65$ & $1.50$ & $2.90$ & $3.10$ \\ 
CS 22945-028 & $4900$ & $1.75$ & $-2.89$ & $1.50$ & $2.50$ & $2.70$ \\ 
CS 22958-074 & $5800$ & $3.60$ & $-2.62$ & $1.40$ & $2.76$ & $2.96$ \\ 
CS 22886-012 & $5650$ & $3.50$ & $-2.61$ & $1.40$ & $2.71$ & $2.92$ \\ 
CS 22885-203 & $5820$ & $3.60$ & $-2.57$ & $1.30$ & $2.71$ & $2.91$ \\ 
CS 22186-002 & $5500$ & $3.35$ & $-2.50$ & $1.10$ & $2.52$ & $2.71$ \\ 
CS 22186-017 & $5770$ & $3.55$ & $-2.90$ & $1.30$ & $2.60$ & $2.79$ \\ 
CS 22942-011 & $4930$ & $1.85$ & $-2.83$ & $1.20$ & $2.36$ & $2.55$ \\ 
CS 29514-018 & $5990$ & $3.70$ & $-2.58$ & $1.20$ & $2.67$ & $2.86$ \\ 
CD -36 1052 & $6030$ & $2.05$ & $-1.86$ & $3.30$ & $3.53$ & $3.72$ \\ 
CS 30339-015 & $5840$ & $3.60$ & $-2.86$ & $1.40$ & $2.53$ & $2.72$ \\ 
CS 22896-015 & $5080$ & $2.25$ & $-2.84$ & $1.20$ & $2.48$ & $2.67$ \\ 
BD +24 1676 & $6140$ & $3.75$ & $-2.54$ & $1.40$ & $2.72$ & $2.91$ \\ 
HE 0938+0114 & $6030$ & $3.65$ & $-2.92$ & $1.20$ & $2.31$ & $2.49$ \\ 
CS 22951-059 & $5120$ & $2.35$ & $-2.83$ & $1.50$ & $2.56$ & $2.74$ \\ 
CS 22183-031 & $4850$ & $1.60$ & $-3.50$ & $1.60$ & $2.30$ & $2.47$ \\ 
CS 22956-106 & $6410$ & $3.90$ & $-2.60$ & $1.80$ & $2.80$ & $2.97$ \\ 
CS 22893-010 & $5150$ & $2.45$ & $-2.93$ & $1.40$ & $2.38$ & $2.55$ \\ 
CS 29502-092 & $4820$ & $1.50$ & $-3.20$ & $1.50$ & $2.06$ & $2.22$ \\ 
G 190-015 & $4950$ & $3.85$ & $-3.13$ & $1.50$ & $2.17$ & $2.31$ \\ 
\end{longtable}
}
\end{appendix}

\end{document}